\documentclass[final,3p,times, onecolumn]{elsarticle}

\usepackage{latexsym}
\usepackage{bm}

\usepackage{amsmath}
\usepackage{graphicx}   
\usepackage{enumerate}   
\usepackage{amssymb}
\begin{document}

\begin{frontmatter}



\title{Introduction to Gauge Theory of Gravitation}

\author{Wytler Cordeiro dos Santos}
\ead{wytler@fis.unb.br}
\address{Universidade de
Bras\'\i lia, CEP 70910-900, DF, Brasil}

\begin{abstract}
The fundamental interactions of nature, the electroweak and the quantum chromodynamics, are described in the Standard Model by the Gauge Theory under internal symmetries that maintain the invariance of the functional action. The fundamental interaction of gravitation is very well described by Einstein's General Relativity in a Riemannian spacetime metric, but General Relativity has been over time a gravitational field theory apart from the Standard Model. The theory of Gauge allows under symmetries of the group of Poincar\'e to impose invariances in the functional of the action of the spinor field that result in the gravitational interaction with the fermions. In this approach the gravitational field, besides being described by the equation similar to General Relativity, also brings a spin-gravitational interaction in a Riemann-Cartan spacetime. 
\end{abstract}


\end{frontmatter}

% \linenumbers

%\label{}

\section{Introduction}

In Classical Mechanics from the Lagrangian formalism it is possible to understand the fundamental laws of Physics. The functional action produces motion equations and under the invariance of the functional action it is possible to obtain the conserved quantities of the motion such as energy and momentum under translational invariance and the angular momentum under invariance by rotations \cite {Goldstein}. From these beautiful concepts of Classical Mechanics, especially Dirac and Feynman, they have shown that the formalism of Lagrange and the action acquire a great and complete importance in the Classical Theory of Fields \cite {Ramond}. The functional action for a field is mathematically described by
\begin{equation}
 S(t_1,t_2,[\Phi]) \equiv \int_{t_1}^{t_2} d^4 \mbox{x}\, {\cal L}(\Phi,\partial_{\alpha}\Phi),
\end{equation}
where  $d^4 \mbox{x} =dt\,dx\, dy\,dz$ 
is a four-dimensional measure of Minkowski spacetime. The integrand ${\cal L}$ is the Lagrangian density of the field function and its derivatives limited by the translational invariance condition. Fields or collections of fields $\Phi$ can be scalar, spinorial, vector, etc. fields. The principle of minimum action results in the Euler-Lagrange equations of motion,
\begin{equation}
\label{Èuler_Lagrange}
 \partial_{\alpha}\left(\frac{\partial{\cal L}}{\partial(\partial_{\alpha}\Phi)}\right)- 
 \frac{\partial{\cal L}} {\partial \Phi}=0,
\end{equation}
which produce (i) the Klein-Gordon equation for the scalar field of sipn zero; (ii) the Dirac equation for the spin field $\frac{1}{2}$; (iii) the Maxwell equations for the electromagnetism, the vector field with spin  1; and so on.

In the Classical Theory of Fields, Lagrangian formalism shows us how to obtain the physical quantities that are conserved, that is, observables that are independent of time.
The mathematical description that
associated a given invariance to a conservation law is the Noether Theorem: if the functional action is invariant under a continuous group of field transformations,
then the Lagrangian establishes a set of dynamic invariants, that is, the conserved currents.

If we consider a continuous group of translation in the spacetime coordinates,
\begin{equation}
 x'^{\alpha} = x^{\alpha}+ a^{\alpha}, 
\end{equation}
in a certain action of relativistic field, we will have as consequence, of the Noether theorem, the conservation of the energy and momentum given by the expression,
\begin{equation}
\partial_{\alpha}T^{\alpha\beta}=0,
\end{equation}
where $T^{\alpha\beta}$ is the energy-momentum tensor. Initially from Lagrangian through the Noether theorem,  we get the canonical energy-momentum tensor given by,
\begin{equation}
\label{energia-momento-canonico}
 {T^{(C) \alpha}}_{\beta} = \left(\frac{\partial{\cal L}}{\partial(\partial_{\alpha}\Phi)}\right)(\partial_{\beta}\Phi) -  {\delta^{\alpha}}_{\beta}{\cal L}.
\end{equation}

For the spin zero scalar field the canonical energy-momentum tensor is symmetric. Since the spin of the scalar field is zero, there are no energy contibutions due to the angular momentum at the energy-momentum tensor. However for the spinorial and vector fields the canonical energy-momentum tensor obtained by (\ref{energia-momento-canonico}) is not symmetrical. The spinorial and vector fields provide energy contribuctions due to the intrinsic angular momentum. The mathematical formalism that fixes this lack of symmetry is Belifante-Rosenfeld's theory. This formalism takes into account the angular momentum of the spinorial and vector fields,
\begin{equation}
\label{tensor_BR_1}
{\mathfrak T}_{\alpha\beta} = T^{(C)}_{\alpha\beta}-\frac{i}{2}\partial^{\gamma}B_{\alpha\beta\gamma},
\end{equation}
where ${\mathfrak T}_{\alpha\beta}$ is the symmetric energy-momentum tensor
and the tensor $B_{\alpha\beta\gamma}$ is given by
\begin{equation}
\label{Belifante_0}
 B_{\alpha\beta\gamma} = S_{\alpha\beta\gamma} + S_{\gamma\alpha\beta} - S_{\beta\gamma\alpha}.
\end{equation}
The tensor $S_{\alpha\beta\gamma}$ is the spin current density tensor contained in the $\Phi$ field. In the formalism of Belifante-Rosenfeld this tensor $S_{\alpha\beta\gamma}$ is given in terms of the Lorentz transformations to the field $\Phi$,
\begin{equation}
 \label{densidade_spin_0}
 {S^{\alpha}}_{\beta\gamma} = \frac{\partial{\cal L}}{\partial(\partial_{\alpha}\Phi)}\Sigma_{\beta\gamma}\,\Phi,
\end{equation}
where the terms $\Sigma_{\beta\gamma}$ are the generators of Lorentz transformations, rotations and boosts, so that the field transforms itself according to Lorentz invariance,
\begin{equation}
\label{transformacao_Lorentz}
 \Phi'(x') = \exp\left[\frac{i}{2}\epsilon^{\alpha\beta}\Sigma_{\alpha\beta}\right]\Phi(x),
\end{equation}
and that obey Lie algebra
\begin{equation}
\label{algebra_Lie_0}
 i[\Sigma_{\gamma\delta}, \Sigma_{\alpha\beta}] = g_{\alpha\delta}\Sigma_{\gamma\beta} -
  g_{\alpha\gamma}\Sigma_{\delta\beta} + g_{\beta\delta}\Sigma_{\alpha\gamma} -g_{\beta\gamma}\Sigma_{\alpha\delta}.
\end{equation}

Field theory for gravitation is very well described by the General Theory of Relavity as a description of a Riemannian spacetime  $({\cal M},{\bm g})$ 
with curvature satisfying Einstein's field equations,
\begin{equation}
\label{equacao_de_campo_Einstein_1}
 G_{\mu\nu} = 8\pi G\, T_{\mu\nu},
\end{equation}
where it is characterized by energy-momentum tensor $T_{\mu\nu}$, 
which accommodates among its components the energy density
and the momentum density associated with a physical system that can curve spacetime. From Noether theorem,
$T_{\mu\nu}$ must satisfy the conservation equation,
\begin{equation}
\nabla_{\mu}T^{\mu\nu}=0.
\end{equation}

In this work we will use the Greek letters: $\alpha,\,\beta,\,\gamma,\,\delta,\cdots$ for the indices of tensors, vectors, etc. in coordinates in the Minkowski spacetime with metric tensor $\eta_{\alpha\beta} = \mbox{diag}(-1,1,1,1)$. Whereas the Greek letters $\kappa,\,\lambda,\,\mu,\,\nu,\cdots $ 
for the indices of tensors, vectors, etc. in the curved spacetime, also denominated holonomic or coordinated indices or yet  world indices. 

As we know the theory of General Relativity describes the trajectories of massive particles and even the photons around large concentrations of masses such as planets, stars, galaxies, etc.
It is reasonable to see the theory of General Relativity as a macroscopic limit of a still unknown quantum theory of gravitation. \'Elie Cartan took a first step toward a possible microscopic theory of gravitation by showing that spacetime torsion could be derived from the spin of matter \cite {Hehl_1}. The initial construction for a theory of gravitation involving spin and torsion was initiated by Utiyama \cite {Utiyama}, and improved by Kibble \cite{Kibble} and Sciama \cite{Sciama} that led to the formulation of the Einstein-Cartan-Kibble-Sciama (ECKS) theory of gravitation \cite{Hehl_1, Hehl_2, Hehl_3, Hehl_4, Pronin}. The ECKS theory of gravitation is described in spacetime called the Riemann-Cartan spacetime denoted by $U_4$, where the spacetime torsion is an integral part of the affine connection. 
In this theory the energy-momentum tensor is coupled with the curvature of spacetime in the same way as in General Relativity as seen in equation (\ref{equacao_de_campo_Einstein_1}), and also produces another equation that couples the spin current density tensor with the $U_4$ spacetime torsion. By coupling energy-momentum and spin of matter to metric and torsion, and by treating them as independent variables, the ECKS theory becomes an extension of General Relativity. The predictions of the ECKS theory are absolutely indistinguishable from the theory of General Relativity even for very high densities of matter, occurring ruptures between the two theories only in ultra high densities. In these extreme environments of ultra densities, the coupling between spin and torsion could in principle produce a gravitational repulsion that would prevent the formation of singularities \cite{Pasmatsiou}. 
In the section \ref{secao_espaco-tempo_de_EC} 
we will see the definitions of torsion and contorsion tensors. In the section \ref{secao_acao_Einstein_Cartan} we will see the main consequences of torsion in the definition of the tensor of curvature, Palatini's identity in spacetime with torsion, the Einstein-Cartan action, and the two Einstein-Cartan field equations.
In the section \ref{campo_espinorial} we make a brief review and discussion on the spinor field, the fundamental blocks of matter in flat spacetime. By the Noether theorem, the energy-momentum tensor is non-symmetric and it is necessary to apply the mechanism of Belifante-Rosenfeld that adds to the canonical energy-momentum tensor the contribution of the spin current density, symmetrizing it.
In the section \ref{secao_Teoria_Gauge_Gravitacao} we present the detailed calculations for the  Gauge Theory of Gravitation  in order to construct the covariant derivative for the spinor action.
And finally in the section \ref{secao_tensor_energia-momento} we show how the energy-momentum tensor and spin current density tensor are obtained from the spinor action in the curved spacetime. From these two tensors the two Einstein-Cartan equations are obtained.

%%%%%%%%%%%%%%%%%%%%%%%%%%%%%%%%%%%%%%%%%%%%%%%%%%%%%%%%%%%%%%

%%%%%%%%%%%%%%%%%%%%%%%%%%%%%%%%%%%%%%%%%%%%%%%%%%%%%%%%%%%%%%

%%%%%%%%%%%%%%%%%%%%%%%%%%%%%%%%%%%%%%%%%%%%%%%%%%%%%%%%%%%%%%

%%%%%%%%%%%%%%%%%%%%%%%%%%%%%%%%%%%%%%%%%%%%%%%%%%%%%%%%%
%%%%%%%%%%%%%%%%%%%%%%%%%%%%%%%%%%%%%%%%%%%%%%%%%%%%%%%%%

\section{Riemann-Cartan spacetime} \label{secao_espaco-tempo_de_EC}

Before we describe the gravitational field equations in a spacetime involving curvature and at the same time the torsion, let us analyze under what conditions the torsion tensor is defined. The starting point for this is that the metric tensor $g_{\mu\nu}$ of a given spacetime has the null covariant derivative,
\begin{equation}
\label{condicao_metrica_1}
\nabla_{\lambda} g_{\mu\nu} \equiv 0.
\end{equation}
The covariant derivative in the Riemann-Cartan spacetime has the same definition and form as the covariant derivative in the Riemannian space,
\begin{equation}
\label{derivada_covariante_1}
\nabla_{\mu} A^{\nu} =\partial_{\mu} A^{\nu} +{\Gamma^{\nu}}_{\mu\kappa}A^{\kappa},
\end{equation}
where the terms ${\Gamma^{\nu}}_{\mu\kappa} $ are metric connection \cite{Nakahara}. 

In the Riemannian spacetime the connections are symmetrical in the indices $\mu$ and $\kappa$,  however in the Riemann-Cartan spacetime the connections are not symmetrical. It is important to note here that the index $\mu$ in the operator 
$\nabla$ of the covariant derivative in the above equation (\ref{derivada_covariante_1}) appears in the second position of the connection coefficient $\Gamma$, whereas some references this index appears in the third position.

With the condition that the covariant derivative of the metric tensor is zero (\ref{condicao_metrica_1}), we can analyze in the same way that one makes in the Riemannian space with a set of three equations seen below,
\begin{equation}
\label{sistema_1}
\begin{cases}
\partial _{\lambda} g_{\mu\nu} - {\Gamma^{\kappa}}_{\lambda\mu} g_{\kappa\nu} - {\Gamma^{\kappa}}_{\lambda\nu} g_{\kappa\mu} = 0 \cr
\partial _{\mu} g_{\nu\lambda} - {\Gamma^{\kappa}}_{\mu\nu} g_{\kappa\lambda} - {\Gamma^{\kappa}}_{\mu\lambda} g_{\kappa\nu} = 0 \cr
\partial _{\nu} g_{\lambda\mu} - {\Gamma^{\kappa}}_{\nu\lambda} g_{\kappa\mu} - {\Gamma^{\kappa}}_{\nu\mu} g_{\kappa\lambda} = 0 .
\end{cases}
\end{equation}
We can perform the following operation with the system of equations above,
\begin{equation}
\label{condicao_metrica_2}
- \partial _{\lambda} g_{\mu\nu} + \partial _{\mu} g_{\nu\lambda} + \partial _{\nu} g_{\lambda\mu}  - \left( {\Gamma^{\kappa}}_{\mu\nu} + {\Gamma^{\kappa}}_{\nu\mu} \right)g_{\kappa\lambda} +  \left({\Gamma^{\kappa}}_{\lambda\mu}  - {\Gamma^{\kappa}}_{\mu\lambda}  \right)g_{\kappa\nu} + \left({\Gamma^{\kappa}}_{\lambda\nu}  - {\Gamma^{\kappa}}_{\nu\lambda}  \right)g_{\kappa\mu} =0.
\end{equation}
The two terms in parenthesis that are subtractions must be defined as torsion tensors,
\begin{equation}
\label{tensor_torcao_1}
{{\mathsf T}^{\kappa}}_{\lambda\mu}  = \left({\Gamma^{\kappa}}_{\lambda\mu}  - {\Gamma^{\kappa}}_{\mu\lambda}  \right).
\end{equation}

In Riemannian spacetime due to symmetry in the indices $\lambda$ and $\mu$ 
the torsion tensor is canceled. Then with the above definition the equation (\ref{condicao_metrica_2}) becomes,
\begin{equation}
- \partial _{\lambda} g_{\mu\nu} + \partial _{\mu} g_{\nu\lambda} + \partial _{\nu} g_{\lambda\mu}  + {{\mathsf T}^{\kappa}}_{\lambda\mu} g_{\kappa\nu} +{{\mathsf T}^{\kappa}}_{\lambda\nu} g_{\kappa\mu}  - \left( {\Gamma^{\kappa}}_{\mu\nu} + {\Gamma^{\kappa}}_{\nu\mu} \right)g_{\kappa\lambda}  = 0, \nonumber
\end{equation}
or yet,
\begin{equation}
 {\mathsf T}_{\mu\lambda\nu} +{\mathsf T}_{\nu\lambda\mu}  +\left(  \partial _{\mu} g_{\nu\lambda} + \partial _{\nu} g_{\lambda\mu}  - \partial _{\lambda} g_{\mu\nu} \right) =  \left( {\Gamma^{\kappa}}_{\mu\nu} + {\Gamma^{\kappa}}_{\nu\mu} \right)g_{\kappa\lambda} . \nonumber
\end{equation}
resulting in
\begin{equation}
\label{conexao_1}
  {\Gamma^{\rho}}_{(\mu\nu)}  =  \frac{1}{2} g^{\lambda\rho} \left(  \partial _{\mu} g_{\nu\lambda} + \partial _{\nu} g_{\lambda\mu}  - \partial _{\lambda} g_{\mu\nu} \right) +\frac{1}{2}g^{\lambda\rho}  \left({\mathsf T}_{\mu\lambda\nu} +{\mathsf T}_{\nu\lambda\mu} \right),
\end{equation}
where we use identity $\frac{1}{2}(A_{\mu}B{\nu}+A_{\nu}B_{\mu}) = A_{(\mu}B_{\nu)} $. 
Notice that the first term on the right-hand side of the above equation is the Christoffel symbol given by
\begin{equation}
\label{Christoffel_1}
\left\{    \begin{matrix} \rho \cr {\mu\nu}  \end{matrix} \right\} =  \frac{1}{2} g^{\lambda\rho} \left(  \partial _{\mu} g_{\nu\lambda} + \partial _{\nu} g_{\lambda\mu}  - \partial _{\lambda} g_{\mu\nu} \right).
\end{equation}

Noting that the Christoffel symbols are symmetric at the $\mu$ and $\nu$ indices, and that the torsion tensor (\ref{tensor_torcao_1}) 
is antisymmetric in the same indices, we can infer that the
$  {\Gamma^{\rho}}_{\mu\nu}  $ 
must be composed of symmetrical and antisymmetric parts,
\begin{equation}
  {\Gamma^{\rho}}_{\mu\nu}  =  {\Gamma^{\rho}}_{(\mu\nu)} + {\Gamma^{\rho}}_{[\mu\nu]} , \nonumber
\end{equation}
where the antisymmetry of the tensor is given by the commutator $ {\Gamma^{\rho}}_{[\mu\nu]} = \frac{1}{2} ( {\Gamma^{\rho}}_{\mu\nu} - {\Gamma^{\rho}}_{\nu\mu}) $, so that we have from the equation above the relation,
\begin{equation}
   {\Gamma^{\rho}}_{(\mu\nu)}  = {\Gamma^{\rho}}_{\mu\nu}  -\frac{1}{2}{\mathsf T^{\rho}}_{\mu\nu}  , \nonumber
\end{equation}
so that replacing this relationship above and the symbol of Christoffel (\ref{Christoffel_1}) in the equation (\ref{conexao_1}) we have
\begin{equation}
  {\Gamma^{\rho}}_{\mu\nu}  =  \left\{    \begin{matrix} \rho \cr {\mu\nu}  \end{matrix} \right\}  + \frac{1}{2}g^{\lambda\rho}  \left({\mathsf T}_{\mu\lambda\nu} +{\mathsf T}_{\nu\lambda\mu} + {\mathsf T}_{\lambda\mu\nu} \right)
\end{equation}
or simply
\begin{equation}
\label{conexao_2}
  {\Gamma^{\rho}}_{\mu\nu}  =  \left\{    \begin{matrix} \rho \cr {\mu\nu}  \end{matrix} \right\}  + {K^{\rho}}_{\mu\nu},
\end{equation}
where we define the contorsion tensor as,
\begin{equation}
\label{contorcao}
 {K^{\rho}}_{\mu\nu}  = \frac{1}{2} \left({{{\mathsf T}_{\mu}}^{\rho}}_{\nu} +{{{\mathsf T}_{\nu}}^{\rho}}_{\mu} + {{\mathsf T}^{\rho}}_{\mu\nu} \right).
\end{equation}

Let us define the torsion vector as a contraction of the tensor (\ref{tensor_torcao_1}) as follows,
\begin{equation}
\label{vetor_torcao}
{\mathsf T}_{\mu} = {\mathsf T^{\nu}}_{\mu\nu} 
\end{equation}
and due to antisymmetry in the indices $\mu$ and $\nu$  of the torsion tensor    (\ref{tensor_torcao_1}) also we have  ${\mathsf T}_{\mu} = - {\mathsf T^{\rho}}_{\rho\mu} $. It is also possible to obtain from the contorsion tensor,
\begin{equation}
{{K_{\rho}}^{\mu}}_{\mu}  = \frac{1}{2} \left({{\mathsf T}_{\mu\rho}}^{\mu} +{{\mathsf T}^{\mu}}_{\rho\mu} + {{{\mathsf T}_{\rho}}^{\mu}}_{\mu} \right) = 
\frac{1}{2} \left({{\mathsf T}^{\mu}}_{\rho\mu}  +{{\mathsf T}^{\mu}}_{\rho\mu} + 0 \right) = {\mathsf T}_{\rho}  .
\end{equation}
Due to antisymmetry it is verified that ${K^{\mu}}_{\mu\nu} = - {\mathsf T}_{\nu} $ and ${K^{\rho}}_{\mu\rho} = 0$.

\subsection{Commutator of covariant derivatives}

In the Riemannian geometry we can obtain the tensor of curvature or Riemann tensor by calculation of the commutator of covariant derivatives. Let us see below what we will get by calculating the commutator of covariant derivatives taking into account that the connections in the spacetime of Riemann-Cartan are composed of symmetrical and antisymmetric part,
\begin{eqnarray}
\left[\nabla_{\mu}, \nabla_{\nu} \right] A^{\rho} &=& \nabla_{\mu} \nabla_{\nu} A^{\rho} - \nabla_{\nu} \nabla_{\mu} A^{\rho}\cr
&=& \partial_{\mu}\partial_{\nu} A^{\rho} + (\partial_{\mu}{\Gamma^{\rho}}_{\nu\sigma})A^{\sigma} + {\Gamma^{\rho}}_{\nu\sigma}\partial_{\mu}A^{\sigma} + {\Gamma^{\rho}}_{\mu\lambda}\partial_{\nu} A^{\lambda}+{\Gamma^{\rho}}_{\mu\lambda}{\Gamma^{\lambda}}_{\nu\sigma}A^{\sigma}  - {\Gamma^{\lambda}}_{\mu\nu}\partial_{\lambda} A^{\rho} - {\Gamma^{\lambda}}_{\mu\nu}{\Gamma^{\rho}}_{\lambda\sigma}A^{\sigma} \cr
& - &  \partial_{\nu}\partial_{\mu} A^{\rho} - (\partial_{\nu}{\Gamma^{\rho}}_{\mu\sigma})A^{\sigma} - {\Gamma^{\rho}}_{\mu\sigma}\partial_{\nu}A^{\sigma} - {\Gamma^{\rho}}_{\nu\lambda}\partial_{\mu} A^{\lambda} - {\Gamma^{\rho}}_{\nu\lambda}{\Gamma^{\lambda}}_{\mu\sigma}A^{\sigma}  + {\Gamma^{\lambda}}_{\nu\mu}\partial_{\lambda} A^{\rho} + {\Gamma^{\lambda}}_{\nu\mu}{\Gamma^{\rho}}_{\lambda\sigma}A^{\sigma} \cr
&=& \left(\partial_{\mu}{\Gamma^{\rho}}_{\nu\sigma} -  \partial_{\nu}{\Gamma^{\rho}}_{\mu\sigma} + {\Gamma^{\rho}}_{\mu\lambda}{\Gamma^{\lambda}}_{\nu\sigma} - {\Gamma^{\rho}}_{\nu\lambda}{\Gamma^{\lambda}}_{\mu\sigma} \right) A^{\sigma} - \left({\Gamma^{\lambda}}_{\mu\nu} -  {\Gamma^{\lambda}}_{\nu\mu}\right)\left(\partial_{\lambda} A^{\rho} + {\Gamma^{\rho}}_{\lambda\sigma}A^{\sigma}\right), \nonumber
\end{eqnarray}
where we have,
\begin{equation}
\label{comutador_derivadas_covariantes}
\left[\nabla_{\mu}, \nabla_{\nu} \right] A^{\rho} = {R^{\rho}}_{\sigma\mu\nu}A^{\sigma} -  {{\mathsf T}^{\lambda}}_{\mu\nu}\nabla_{\lambda}A^{\rho}.
\end{equation}
Note that if spacetime has no torsion, $ {{\mathsf T}^{\lambda}}_{\mu\nu} = 0 $, 
the above commutator is reduced to the Riemannian space commutator.

\section{The Einstein-Cartan action} \label{secao_acao_Einstein_Cartan}

We have seen in the previous section that the Riemann-Cartan spacetime curvature tensor has the same algebraic format as the Riemann spacetime curvature tensor, with the detail that the Riemann-Cartan spacetime connections have components symmetrical and antisymmetric components,
\begin{equation}
 {R^{\kappa}}_{\lambda\mu\nu} = \partial_{\mu}{\Gamma^{\kappa}}_{\nu\lambda} -  \partial_{\nu}{\Gamma^{\kappa}}_{\mu\lambda} + {\Gamma^{\kappa}}_{\mu\rho}{\Gamma^{\rho}}_{\nu\lambda} -  {\Gamma^{\kappa}}_{\nu\rho}{\Gamma^{\rho}}_{\mu\lambda}.
\end{equation}
An infinitesimal variation in the curvature tensor results in,
\begin{equation}
 \delta {R^{\kappa}}_{\lambda\mu\nu} = \delta\left( \partial_{\mu}{\Gamma^{\kappa}}_{\nu\lambda} \right) -  \delta\left( \partial_{\nu}{\Gamma^{\kappa}}_{\mu\lambda} \right)+  \delta\left(  {\Gamma^{\kappa}}_{\mu\rho}\right) {\Gamma^{\rho}}_{\nu\lambda} +  {\Gamma^{\kappa}}_{\mu\rho} \, \delta {\Gamma^{\rho}}_{\nu\lambda}  -   \delta\left(  {\Gamma^{\kappa}}_{\nu\rho}\right) {\Gamma^{\rho}}_{\mu\lambda} - {\Gamma^{\kappa}}_{\nu\rho}\, \delta {\Gamma^{\rho}}_{\mu\lambda},
\end{equation}
with the condition 
\begin{equation}
\delta\left( \partial_{\mu}{\Gamma^{\kappa}}_{\nu\lambda} \right)  = \partial_{\mu}\, \delta {\Gamma^{\kappa}}_{\nu\lambda}, \nonumber 
\end{equation}
we then obtain that an infinitesimal variation in the tensor of curvature results in,
\begin{equation}
\label{variacao_tensor_curvatura}
 \delta {R^{\kappa}}_{\lambda\mu\nu} = \partial_{\mu}\, \delta {\Gamma^{\kappa}}_{\nu\lambda}  -   \partial_{\nu}\,\delta{\Gamma^{\kappa}}_{\mu\lambda} +  \delta\left(  {\Gamma^{\kappa}}_{\mu\rho}\right) {\Gamma^{\rho}}_{\nu\lambda} +  {\Gamma^{\kappa}}_{\mu\rho} \, \delta {\Gamma^{\rho}}_{\nu\lambda}  -   \delta\left(  {\Gamma^{\kappa}}_{\nu\rho}\right) {\Gamma^{\rho}}_{\mu\lambda} - {\Gamma^{\kappa}}_{\nu\rho}\, \delta {\Gamma^{\rho}}_{\mu\lambda},
\end{equation}
With this equation in mind we will now describe the famous Palatini identity in this context of Riemann-Cartan spacetime given by,
\begin{equation}
\label{palatini_1}
\nabla_{\mu}\left( \delta  {\Gamma^{\kappa}}_{\nu\lambda}  \right) - \nabla_{\nu}\left( \delta  {\Gamma^{\kappa}}_{\mu\lambda}  \right) = \partial_{\mu}\, \delta {\Gamma^{\kappa}}_{\nu\lambda}+ {\Gamma^{\kappa}}_{\mu\rho}\delta {\Gamma^{\rho}}_{\nu\lambda} -{\Gamma^{\rho}}_{\mu\nu}\delta {\Gamma^{\kappa}}_{\rho\lambda}-{\Gamma^{\rho}}_{\mu\lambda}\delta {\Gamma^{\kappa}}_{\nu\rho}
  -  \left[ \partial_{\nu}\,\delta{\Gamma^{\kappa}}_{\mu\lambda} + {\Gamma^{\kappa}}_{\nu\rho}\delta {\Gamma^{\rho}}_{\mu\lambda} -{\Gamma^{\rho}}_{\nu\mu}\delta {\Gamma^{\kappa}}_{\rho\lambda}-{\Gamma^{\rho}}_{\nu\lambda}\delta {\Gamma^{\kappa}}_{\mu\rho}  \right],
\end{equation}
then comparing the above equation (\ref{palatini_1}) 
with the antecedent equation (\ref{variacao_tensor_curvatura}) we have that 
\begin{equation}
\nabla_{\mu}\left( \delta  {\Gamma^{\kappa}}_{\nu\lambda}  \right) - \nabla_{\nu}\left( \delta  {\Gamma^{\kappa}}_{\mu\lambda}  \right) =  \delta {R^{\kappa}}_{\lambda\mu\nu}  - \left( {\Gamma^{\rho}}_{\mu\nu} -  {\Gamma^{\rho}}_{\nu\mu} \right) \delta {\Gamma^{\kappa}}_{\rho\lambda}, \nonumber
\end{equation}
by the definition of the torsion tensor (\ref{tensor_torcao_1}), 
we obtain the identity of Palatini in the spacetime of Riemann-Cartan,
\begin{equation}
\label{palatini_2}
  \delta {R^{\kappa}}_{\lambda\mu\nu}   = \nabla_{\mu}\left( \delta  {\Gamma^{\kappa}}_{\nu\lambda}  \right) - \nabla_{\nu}\left( \delta  {\Gamma^{\kappa}}_{\mu\lambda}  \right) + {{\mathsf T}^{\rho}}_{\mu\nu}  \delta {\Gamma^{\kappa}}_{\rho\lambda}.
\end{equation}
For the gravitational field action it is necessary obtain the Ricci curvature tensor ${R^{\mu}}_{\lambda\mu\nu} =R_{\lambda\nu}$, 
whose infinitesimal variation is given by the equation below,
\begin{equation}
\label{palatini_3}
  \delta R_{\lambda\nu}  = \nabla_{\mu}\left( \delta  {\Gamma^{\mu}}_{\nu\lambda}  \right) - \nabla_{\nu}\left( \delta  {\Gamma^{\mu}}_{\mu\lambda}  \right) + {{\mathsf T}^{\rho}}_{\mu\nu}  \delta {\Gamma^{\mu}}_{\rho\lambda}.
\end{equation}

It is now possible to describe the gravitational field action in Riemann-Cartan spacetime. The action is algebraically similar to the action of Einstein-Hilbert in the General Relavity. In spacetime with torsion this action is called Einstein-Cartan action given by the equation below,
\begin{equation}
\label{acao_EC_1}
S_{EC} =\frac{1}{16\pi G} \int_{\Omega} d^4 x \sqrt{-g}\, R ,
\end{equation}
where $R= g^{\mu\nu}R_{\mu\nu}$ is the scalar curvature. As in General Relativity, the field equations are obtained from the variation of the action,
\begin{equation}
\delta S_{EC} =\frac{1}{16\pi G} \int_{\Omega} d^4 x \left[ \left(\delta \sqrt{-g}\right) \, R +  \sqrt{-g} \left(\delta R_{\mu\nu} \right)g^{\mu\nu} +  \sqrt{-g} R_{\mu\nu} \delta g^{\mu\nu} \right]. \nonumber
\end{equation}
We can substitute in the above equation $\delta\sqrt{-g} = -\frac{\sqrt{-g}}{2} g_{\mu\nu}\delta g^{\mu\nu}$ 
and also the identity (\ref{palatini_3}) 
to obtain
\begin{equation}
\delta S_{EC} =\frac{1}{16\pi G} \int_{\Omega} d^4 x \Bigg\{ -\frac{\sqrt{-g}}{2} g_{\mu\nu}\, R \delta g^{\mu\nu} +  \sqrt{-g} R_{\mu\nu} \delta g^{\mu\nu} 
 +   \sqrt{-g} \left[  \nabla_{\lambda}\left( \delta  {\Gamma^{\lambda}}_{\nu\mu}  \right) - \nabla_{\nu}\left( \delta  {\Gamma^{\lambda}}_{\lambda\mu}  \right) + {{\mathsf T}^{\rho}}_{\kappa\nu}  \delta {\Gamma^{\kappa}}_{\rho\mu}\right]g^{\mu\nu} \Bigg\}. \nonumber
\end{equation}
From the covariance condition of the metric tensor (\ref{condicao_metrica_1}) 
we can obtain
\begin{equation}
\label{acao_EC_2}
\delta S_{EC} = \frac{1}{16\pi G} \int_{\Omega} d^4 x \sqrt{-g} \left[R_{\mu\nu}-\frac{1}{2} g_{\mu\nu}\, R \right]  \delta g^{\mu\nu} 
 +   \frac{1}{16\pi G} \int_{\Omega} d^4 x \sqrt{-g}  \left[  \nabla_{\lambda}\left( g^{\mu\nu} \delta  {\Gamma^{\lambda}}_{\nu\mu}  \right) - \nabla_{\nu}\left( g^{\mu\nu}\delta  {\Gamma^{\lambda}}_{\lambda\mu}  \right) + {{{\mathsf T}^{\rho}}_{\kappa}}^{\mu}  \delta {\Gamma^{\kappa}}_{\rho\mu}\right]. 
\end{equation}

We can already identify Einstein's tensor, $G_{\mu\nu} = R_{\mu\nu}-\frac{1}{2}g_{\mu\nu}\, R$, 
in the first integral of the above equation. The two terms in the second part, $ \nabla_{\lambda}\left( g^{\mu\nu} \delta  {\Gamma^{\lambda}}_{\nu\mu}  \right) - \nabla_{\nu}\left( g^{\mu\nu}\delta  {\Gamma^{\lambda}}_{\lambda\mu}  \right) $ are  divergences, $\nabla_{\mu} V^{\mu}$. 
Let us see in some detail how one should treat these two divergences terms in spacetime with torsion non-zero. First we see that,
\begin{equation}
\nabla_{\mu} V^{\mu} = \partial_{\mu} V^{\mu} + {\Gamma^{\mu}}_{\mu\lambda} V^{\lambda}, \nonumber
\end{equation}
or else,
\begin{equation}
\nabla_{\mu} V^{\mu} = \partial_{\mu} V^{\mu} +   \left\{    \begin{matrix} \mu \cr {\mu\lambda}  \end{matrix} \right\}  V^{\lambda} +{K^{\mu}}_{\mu\lambda} V^{\lambda}, \nonumber
\end{equation}
where we can use the identity ${K^{\mu}}_{\mu\lambda} = - {\mathsf T}_{\lambda}$, 
as well as the identity of Christoffel's symbols,
\begin{equation}
  \left\{    \begin{matrix} \mu \cr {\mu\lambda}  \end{matrix} \right\}  = \frac{\partial_{\lambda}\sqrt{-g}}{\sqrt{-g}}, \nonumber
\end{equation}
we obtain for the divergence of the vector $V^{\mu}$ 
the following equation,
\begin{equation}
\label{divergencia_1}
\nabla_{\mu} V^{\mu} = \partial_{\mu} V^{\mu} + \frac{\partial_{\mu}\sqrt{-g}}{\sqrt{-g}}\, V^{\mu} -  {\mathsf T}_{\mu}\,V^{\mu}.
\end{equation}
If we realize the integral of this divergence in the region $\Omega$,
\begin{equation}
\int_{\Omega} d^4x \sqrt{-g}\, \nabla_{\mu} V^{\mu} = \int_{\Omega} d^4x \partial_{\mu}\left (\sqrt{-g}\, V^{\mu}\right) - \int_{\Omega} d^4x \sqrt{-g}\, {\mathsf T}_{\mu} V^{\mu} ,
\end{equation}
where we can use the Gauss divergence theorem, so that the integral can be rewritten as,
\begin{equation}
\int_{\Omega} d^4x \sqrt{-g}\, \nabla_{\mu} V^{\mu} = \oint_{\partial \Omega} d^3x \sqrt{-g}\, V^{\mu}\hat{n}_{\mu} - \int_{\Omega} d^4x \sqrt{-g}\, {\mathsf T}_{\mu} V^{\mu} .
\end{equation}
In the Classical Theory of Fields, the surface term $\partial\Omega$ 
must be canceled due to finitude of the fields on the boundary \cite{Ramond}, then we have as a result of this integral the value
\begin{equation}
\label{divergencia_2}
\int_{\Omega} d^4x \sqrt{-g}\, \nabla_{\mu} V^{\mu} = - \int_{\Omega} d^4x \sqrt{-g}\, {\mathsf T}_{\mu} V^{\mu} .
\end{equation}

We should note that if we were dealing with a spacetime free of torsion the second integral on the right side of equation (\ref{acao_EC_2}) 
would cancel out and obtain the gravitational field equation of the General Relavity only. Returning the result obtained in equation (\ref{divergencia_2}) 
in the variation of the equation (\ref{acao_EC_2}), 
we get, then,
\begin{eqnarray}
\label{acao_EC_3}
\delta S_{EC} = \frac{1}{16\pi G} \int_{\Omega} d^4 x \sqrt{-g}\,\, G_{\mu\nu} \delta g^{\mu\nu}  +   \frac{1}{16\pi G} \int_{\Omega} d^4 x \sqrt{-g}  \left[  -{\mathsf T}_{\lambda}\left( g^{\mu\nu} \delta  {\Gamma^{\lambda}}_{\nu\mu}  \right)  + {\mathsf T}_{\nu}\left( g^{\mu\nu}\delta  {\Gamma^{\lambda}}_{\lambda\mu}  \right) + {{{\mathsf T}^{\lambda}}_{\mu}}^{\nu}  \delta {\Gamma^{\mu}}_{\lambda\nu}\right]. 
\end{eqnarray}
Before we move on we will use a mathematical identity of the metric tensor that can be seen below as,
\begin{equation}
 \delta g^{\mu\nu} = -g^{\mu\kappa}g^{\nu\lambda}\delta g_{\lambda\kappa},
\end{equation}
and substituting this equality in the variation of the action (\ref{acao_EC_3}) we obtain,
\begin{eqnarray}
\label{acao_EC_4}
\delta S_{EC} = -\frac{1}{16\pi G} \int_{\Omega} d^4x  \sqrt{-g}\,\, G^{\kappa\lambda} \delta g_{\kappa\lambda}  +   \frac{1}{16\pi G} \int_{\Omega} d^4x \sqrt{-g}  \left[  -{\mathsf T}_{\lambda}\left( g^{\mu\nu} \delta  {\Gamma^{\lambda}}_{\nu\mu}  \right)  + {\mathsf T}_{\nu}\left( g^{\mu\nu}\delta  {\Gamma^{\lambda}}_{\lambda\mu}  \right) + {{{\mathsf T}^{\lambda}}_{\mu}}^{\nu}  \delta {\Gamma^{\mu}}_{\lambda\nu}\right]. 
\end{eqnarray}
So finally we get to the point where it is possible to get two field equations from the above functional action. The first is the variation of the action in relation to the metric tensor, which results in the traditional gravitational field equation of General Relativity,
\begin{equation}
\label{equacao_de_campo_Einstein}
\frac{\delta S_{EC}}{\delta g_{\mu\nu}} = -\frac{1}{16\pi G} \, G^{\mu\nu} = -\frac{1}{16\pi G}\left(R^{\mu\nu}-\frac{1}{2} g^{\mu\nu}\, R \right).
\end{equation}
The second variation is in relation to the field of connections leading to,
\begin{equation}
\frac{\delta S_{EC}}{\delta {\Gamma^{\rho}}_{\sigma\tau}} =  \frac{1}{16\pi G}\left(  {{{\mathsf T}^{\lambda}}_{\mu}}^{\nu}  {\delta_{\rho}}^{\mu}  {\delta_{\lambda}}^{\sigma}   {\delta_{\nu}}^{\tau} +  {\mathsf T}_{\nu} g^{\mu\nu} {\delta_{\rho}}^{\lambda}  {\delta_{\lambda}}^{\sigma}   {\delta_{\mu}}^{\tau}  - {\mathsf T}_{\lambda}g^{\mu\nu} {\delta_{\rho}}^{\lambda}  {\delta_{\nu}}^{\sigma}   {\delta_{\mu}}^{\tau}   \right)
= \frac{1}{16\pi G}\left(  {{{\mathsf T}^{\sigma}}_{\rho}}^{\tau}  +  {\mathsf T}_{\nu} g^{\tau\nu} {\delta_{\rho}}^{\sigma}  - {\mathsf T}_{\rho}g^{\tau\sigma} \right)\nonumber
\end{equation}
or else
\begin{equation}
\label{equacao_de_campo_Cartan}
\frac{\delta S_{EC}}{\delta \Gamma^{\rho\sigma\tau}}  =   \frac{1}{16\pi G}\left(  {\mathsf T}_{\sigma\rho\tau}  +  {\mathsf T}_{\tau}g_{\rho\sigma}  - {\mathsf T}_{\rho}g_{\tau\sigma} \right).
\end{equation}

Now consider the presence of a field of matter in Riemann-Cartan spacetime where the total action of a given system is given by
\begin{equation}
S = S_{EC} +S_{M},
\end{equation}
where $S_{M}$ 
is an action of some field of matter given by
\begin{equation}
 S_{M} = \int d^4x \,\,{\cal L}_{M},
\end{equation}
where ${\cal L}_{M}$ 
is a Lagrangian of some field.
Thus, a variation in action results in
\begin{equation}
\delta S_{EC} +\delta S_{M} = 0,
\end{equation}
with 
$$ \frac{\delta S_{EC} }{\delta g_{\mu\nu}}=-\frac{1}{16\pi G}\sqrt{-g}\,\,G^{\mu\nu},$$
thus,
$$ -\frac{\sqrt{-g}}{16\pi G}G^{\mu\nu}+\frac{\delta S_{M} }{\delta g_{\mu\nu}}= 0,$$
or then,
\begin{equation}
G^{\mu\nu}= \frac{16\pi G}{\sqrt{-g}}\frac{\delta S_{M} }{\delta g_{\mu\nu}}.
\end{equation}
Then we can calculate the energy-momentum tensor from the action of a field of matter, so that it results in the Einstein field equation,
\begin{equation}
\label{Tensor_E-P_1}
T^{\mu\nu} = \frac{2}{\sqrt{-g}}\frac{\delta S_{M} }{\delta g_{\mu\nu}}.
\end{equation}

The energy-momentum tensor calculated by the above equation is called the metric energy-momentum tensor or Hilbert energy-momentum tensor. For an action $S_{M}$ 
scalar field and also for the electromagnetic field the above calculation results directly in a symmetric energy-momentum tensor. However for a spin field the energy-momentum tensor calculated by equation (\ref{Tensor_E-P_1}) 
will not be symmetrical, being necessary to use the mechanism of Belinfante \cite{Belifante}. 
In the next section we will show the calculations of this tensor in the flat spacetime of the Special Relativity and later we will show how to calculate this tensor in the curved spacetime.

Let us now look at the result we get if the action variation is in relation to the connection field given by equation (\ref{equacao_de_campo_Cartan}), thus we have that
\begin{equation}
 \frac{\delta S_{EC}}{\delta \Gamma^{\lambda\mu\nu}} +  \frac{\delta S_{M}}{\delta \Gamma^{\lambda\mu\nu}} =   0, \nonumber
\end{equation}
resulting in
\begin{equation}
\label{equacao_de_campo_Cartan_2}
  {\mathsf T}_{\mu\lambda\nu}  +  {\mathsf T}_{\nu}g_{\lambda\mu}  - {\mathsf T}_{\lambda}g_{\mu\nu} =  8\pi G \,\mathfrak{S}_{\mu\lambda\nu},
\end{equation}
where 
\begin{equation}
\label{equacao_de_campo_Cartan_3}
 \mathfrak{S}_{\mu\lambda\nu} =  -\frac{2}{\sqrt -g} \frac{\delta S_{M}}{\delta \Gamma^{\lambda\mu\nu}}
\end{equation}
 is the spin current density tensor of the matter field. The energy-momentum tensor is the source of the curvature in spacetime whereas the spin current density tensor is the source of torsion in spacetime. We will look at this in more detail in the next sections.

%%%%%%%%%%%%%%%%%%%%%%%%%%%%%%%%%%%%%%%%%%%%%%%%%%%%%%%%%%
%%%%%%%%%%%%%%%%%%%%%%%%%%%%%%%%%%%%%%%%%%%%%%%%%%%%%%%%%%
%%%%%%%%%%%%%%%%%%%%%%%%%%%%%%%%%%%%%%%%%%%%%%%%%%%%%%%%%%

\section{Spinorial field}\label{campo_espinorial}

The basic components of matter are the fermions, and the theory of gauge for gravitation must be constructed from the fields of spin fermionic matter $\frac{1}{2}$.
The functional action has Lagrangian in flat Minkowski spacetime given by,
\begin{equation}
 {\cal L}= \frac{i}{2} \left(\bar\psi\gamma^{\alpha}\partial_{\alpha}\psi - 
 (\partial_{\alpha}\bar\psi)\gamma^{\alpha}\psi\right)
 +m\bar\psi \psi.
\end{equation}
Where the  $\gamma^{\alpha}$  matrices obey Clifford's algebra
\begin{equation}
\label{Clifford}
 \{\gamma^{\alpha},\gamma^{\beta}\}= \gamma^{\alpha}\gamma^{\beta}+ \gamma^{\beta}\gamma^{\alpha}=2\eta^{\alpha\beta} 1\!\!1.
\end{equation}
where $1\!\!1$ is the identity $4\times 4$ matrix.
The fermionic field $\psi$ 
is a four-component column spinor, whose adjoint spinor is given by
$\bar\psi=\psi^{\dag}\gamma^0$.

The Euler-Lagrange equations for the above spinorial Lagrangian are the two equations below 
\begin{equation}
 \partial_{\alpha}\left(\frac{\partial{\cal L}}{\partial(\partial_{\alpha}\bar\psi)}\right)- 
 \frac{\partial{\cal L}} {\partial \bar\psi}=0.
\end{equation}
and
\begin{equation}
 \partial_{\alpha}\left(\frac{\partial{\cal L}}{\partial(\partial_{\alpha}\psi)}\right)- 
 \frac{\partial{\cal L}} {\partial \psi}=0.
\end{equation}
which result in the two Dirac equations
\begin{equation}
\label{Dirac_1}
 (i\gamma^{\alpha}\partial_{\alpha} + m 1\!\!1)\psi=0
\end{equation}
and 
\begin{equation}
\label{Dirac_2}
  -i\partial_{\alpha}\bar\psi\gamma^{\alpha} +m\bar\psi = 0 \hspace*{1cm}\mbox{or} \hspace*{1cm} \bar\psi(-i\overleftarrow{\partial}_{\alpha}\gamma^{\alpha} + m 1\!\!1)=0.
\end{equation}

The Noether theorem shows us that the canonical energy-momentum tensor will be given by
\begin{equation}
\label{canonico_1}
 {T^{(C)\alpha}}_{\beta}= \frac{\partial{\cal L}}{\partial(\partial_{\alpha}\psi)}(\partial_{\beta}\psi)+
\partial_{\beta}\bar\psi\frac{\partial{\cal L}}{\partial(\partial_{\alpha}\bar\psi)} - {\delta^{\alpha}}_{\beta}{\cal L},
\end{equation}
which results in
\begin{equation}
\label{canonico_2}
 T^{(C)}_{\alpha\beta} =\frac{i}{2}\bar\psi\gamma_{\alpha}\partial_{\beta}\psi -
 \frac{i}{2}(\partial_{\beta}\bar\psi)\gamma_{\alpha}\psi.
\end{equation}

It should be noted here that the canonical energy-momentum tensor obtained through Noether theorem is not symmetric. Therefore, it is necessary to use the Belinfante-Rosenfeld procedures didatically discussed by Weinberg in reference \cite{Weinberg}. 
The Belinfante-Rosenfeld methodology consists in calculating the energy density contained in spin density. This methodology is summarized in calculating the equation (\ref{tensor_BR_1}),
\begin{equation}
\label{tensor_BR}
 {\mathfrak T}_{\alpha\beta} = T^{(C)}_{\alpha\beta}-\frac{i}{2}\partial^{\delta}B_{\alpha\beta\delta},
\end{equation}
or
\begin{equation}
\label{tensor_Belinfante-Rosenfeld}
 {\mathfrak T}_{\alpha\beta} = T^{(C)}_{\alpha\beta}-\frac{i}{2}\partial^{\delta}(S_{\alpha\beta\delta} + S_{\delta\alpha\beta} - S_{\beta\delta\alpha}),
\end{equation}
where ${\mathfrak T}_{\alpha\beta}$ is the symmetric energy-momentum tensor and the tensor $B_{\alpha\beta\delta}$ is given by equation (\ref{Belifante_0}) as follows
\begin{equation}
\label{Belifante_1}
 B_{\alpha\beta\delta} = S_{\alpha\beta\delta} + S_{\delta\alpha\beta} - S_{\beta\delta\alpha}.
\end{equation}
The tensor $S_{\alpha\beta\delta}$ is the density of spin contained in the fermionic field that according to the Classical Theory of Fields is given by,
\begin{equation}
 \label{densidade_spin_1}
 {S^{\alpha}}_{\beta\delta} = \frac{\partial{\cal L}}{\partial(\partial_{\alpha}\psi)}\Sigma_{\beta\delta}\,\psi + \bar{\psi}(-\Sigma_{\beta\delta})\frac{\partial{\cal L}}{\partial(\partial_{\alpha}\bar{\psi})},
\end{equation}
reaffirming that the adjoint spinor $\bar\psi$ 
is transformed with a signal exchanged in relation to $\psi$, so the negative signal in the generator of the Lorentz transformations  $\Sigma_{\beta\delta}$ for spinorial field. By performing this calculation we obtain that
\begin{equation}
 \label{densidade_spin_2}
 {S^{\alpha}}_{\beta\delta} = \frac{i}{2}\bar\psi \gamma^{\alpha}\Sigma_{\beta\delta}\psi - \bar\psi\Sigma_{\beta\delta}\left(\frac{-i}{2}\right)\gamma^{\alpha}\psi = \frac{i}{2} \bar\psi \left(\gamma^{\alpha}\Sigma_{\beta\delta}+\Sigma_{\beta\delta}\gamma^{\alpha}\right)\psi,
\end{equation}
or in terms of anticommutator,
\begin{equation}
 \label{densidade_spin_3}
 S_{\alpha\beta\delta} = \frac{i}{2}\,\bar\psi \left\{\gamma_{\alpha},\Sigma_{\beta\delta}\right\}\psi.
\end{equation}
Now we can calculate the tensor $B_{\alpha\beta\delta}$ in the equation (\ref{Belifante_1}), where we have
\begin{equation}
\label{Belifante_2}
 B_{\alpha\beta\delta} = S_{\alpha\beta\delta} + S_{\delta\alpha\beta} - S_{\beta\delta\alpha} = \frac{i}{2}\,\bar\psi \left(\left\{\gamma_{\alpha},\Sigma_{\beta\delta}\right\} + \left\{\gamma_{\delta},\Sigma_{\alpha\beta}\right\} -  \left\{\gamma_{\beta},\Sigma_{\delta\alpha}\right\}\right)\psi.
\end{equation}
We can now simplify the above expression using the definition of Lorentz transformation generators,
\begin{equation}
\label{gerador_Lorentz}
 \Sigma_{\alpha\beta} = \frac{i}{4}\left[ \gamma_{\alpha},\gamma_{\beta}\right],
\end{equation}
also the identity (\ref{Clifford}), where $\{\gamma^{\alpha},\gamma^{\beta}\}= 2\eta^{\alpha\beta} 1\!\!1$ together with the identity
\begin{equation}
 \left\{A,\left[B,C\right]\right\} - \left\{B,\left[C, A\right]\right\} = \left[\left\{A,B\right\}, C\right],
\end{equation}
so that the last two anticommutator in the expression (\ref{Belifante_2}) will be given by
\begin{equation}
 \left\{\gamma_{\beta},\Sigma_{\delta\alpha}\right\} - \left\{\gamma_{\delta},\Sigma_{\alpha\beta}\right\} =  \frac{i}{4}\left\{\gamma_{\beta},[\gamma_{\delta},\gamma_{\alpha}]\right\} - \frac{i}{4}\left\{\gamma_{\delta},[\gamma_{\alpha},\gamma_{\beta}]\right\} =\frac{i}{4}\left[\left\{\gamma_{\beta},\gamma_{\delta}\right\},\gamma_{\alpha}\right] = \frac{i}{4}\left[2\eta_{\beta\delta}1\!\!1 ,\gamma_{\alpha}\right] = 0.
\end{equation}
Then the tensor of equation (\ref{Belifante_2}) results in
\begin{equation}
\label{Belifante_3}
 B_{\alpha\beta\delta}  = \frac{i}{2}\,\bar\psi \left\{\gamma_{\alpha},\Sigma_{\beta\delta}\right\}\psi.
\end{equation}
To finalize the calculation of the energy-momentum tensor of Belifante-Rosenfeld in the equation (\ref{tensor_BR}), 
we will calculate the derivative of the tensor above
\begin{equation}
\label{Belifante_4}
 \partial^{\delta} B_{\alpha\beta\delta}  = \frac{i}{2}\,(\partial^{\delta}\bar\psi) \left\{\gamma_{\alpha},\Sigma_{\beta\delta}\right\}\psi + \frac{i}{2}\,\bar\psi \left\{\gamma_{\alpha},\Sigma_{\beta\delta}\right\}\partial^{\delta}\psi.
\end{equation}
Now we must see the calculations in the above terms that can be simplified. For the first term in the above equation commutating the $\gamma_{\delta}$ matrix  
to the left, we have
\begin{equation}
 \frac{i}{2}\,(\partial^{\delta}\bar\psi) \left\{\gamma_{\alpha},\Sigma_{\beta\delta}\right\}\psi =  \frac{i}{2}\,(\partial^{\delta}\bar\psi)\frac{i}{4}\left(4\eta_{\beta\delta}\gamma_{\alpha} - 4\eta_{\alpha\delta}\gamma_{\beta}+2\gamma_{\delta}[\gamma_{\alpha},\gamma_{\beta}] \right)\psi
 = \frac{i^2}{2}\left( (\partial_{\beta}\bar\psi)\gamma_{\alpha} -(\partial_{\alpha}\bar\psi)\gamma_{\beta}+\frac{1}{2}\underbrace{ (\partial^{\delta}\bar\psi)\gamma_{\delta}}[\gamma_{\alpha},\gamma_{\beta}] \right)\psi, \nonumber
\end{equation}
where we must use the Dirac equation (\ref{Dirac_2}) 
in the above underbraced term resulting in
\begin{equation}
\label{Belifante_5}
 \frac{i}{2}\,(\partial^{\delta}\bar\psi) \left\{\gamma_{\alpha},\Sigma_{\beta\delta}\right\}\psi = -\frac{1}{2} \left((\partial_{\beta}\bar\psi)\gamma_{\alpha} -(\partial_{\alpha}\bar\psi)\gamma_{\beta}-\frac{i}{2}m\bar\psi[\gamma_{\alpha},\gamma_{\beta}]  \right)\psi.
\end{equation}
For the second term of the equation (\ref{Belifante_4}) let us commute $\gamma_{\delta}$ matrix  to the right side,
\begin{equation}
  \frac{i}{2}\,\bar\psi \left\{\gamma_{\alpha},\Sigma_{\beta\delta}\right\}\partial^{\delta}\psi = \frac{i}{2}\,\bar\psi\frac{i}{4}\left( 4\eta_{\delta\alpha}\gamma_{\beta} -4\eta_{\beta\delta}\gamma_{\alpha}+2[\gamma_{\alpha},\gamma_{\beta}]\gamma_{\delta}\right)\partial^{\delta}\psi
  = \frac{i^2}{2}\left( \bar\psi \gamma_{\beta}\partial_{\alpha} \psi - \bar\psi \gamma_{\alpha}\partial_{\beta} \psi + \frac{1}{2}\bar\psi[\gamma_{\alpha},\gamma_{\beta}]\underbrace{\gamma_{\delta}\partial^{\delta}\psi}\right),\nonumber
\end{equation}
where we must use the Dirac equation (\ref{Dirac_1}) 
in the the above underbraced term resulting in,
\begin{equation}
\label{Belifante_6}
  \frac{i}{2}\,\bar\psi \left\{\gamma_{\alpha},\Sigma_{\beta\delta}\right\}\partial^{\delta}\psi = -\frac{1}{2}\left( \bar\psi \gamma_{\beta}\partial_{\alpha} \psi - \bar\psi \gamma_{\alpha}\partial_{\beta} \psi + \frac{i}{2}m\bar\psi[\gamma_{\alpha},\gamma_{\beta}]\psi\right).
\end{equation}
Then adding the two results (\ref{Belifante_5}) and (\ref{Belifante_6}) to obtain the result for the equation (\ref{Belifante_4}),
\begin{equation}
\label{Belifante_7}
 \partial^{\delta} B_{\alpha\beta\delta}  = -\frac{1}{2} \left((\partial_{\beta}\bar\psi)\gamma_{\alpha}\psi -(\partial_{\alpha}\bar\psi)\gamma_{\beta}\psi + \bar\psi \gamma_{\beta}\partial_{\alpha} \psi - \bar\psi \gamma_{\alpha}\partial_{\beta} \psi\right).
\end{equation}
So we can finally put this result into the expression (\ref{tensor_BR}) to obtain the Belifante-Rosenfeld symmetrized energy-momentum tensor of the  spinor field
\begin{equation}
 {\mathfrak T}_{\alpha\beta} = T^{(C)}_{\alpha\beta}-\frac{i}{2}\partial^{\delta}B_{\alpha\beta\delta}
 = \frac{i}{2}\bar\psi\gamma_{\alpha}\partial_{\beta}\psi -
 \frac{i}{2}(\partial_{\beta}\bar\psi)\gamma_{\alpha}\psi + \frac{i}{4} \left((\partial_{\beta}\bar\psi)\gamma_{\alpha}\psi -(\partial_{\alpha}\bar\psi)\gamma_{\beta}\psi + \bar\psi \gamma_{\beta}\partial_{\alpha} \psi - \bar\psi \gamma_{\alpha}\partial_{\beta} \psi\right)
 ,\nonumber
\end{equation}
where we got to the result
\begin{equation}
\label{tensor_energia_momento}
 {\mathfrak T}_{\alpha\beta} =\frac{i}{4}(\bar\psi\gamma_{\alpha}\partial_{\beta}\psi + \bar\psi\gamma_{\beta}\partial_{\alpha}\psi)
 - \frac{i}{4}( (\partial_{\alpha}\bar\psi)\gamma_{\beta}\psi+(\partial_{\beta}\bar\psi)\gamma_{\alpha}\psi),
\end{equation}
in which the energy-momentum tensor of the spinorial field is finally symmetrized.

The transition from a physical equation in Minkowski flat spacetime  to a curved spacetime is done by replacing the metric tensor $\eta_{\alpha\beta}$ from flat spacetime by metric tensor $g_{\mu\nu}$ of a curved spacetime. 
The principle of equivalence tells us that we must replace the derivatives $\partial_{\alpha}$ from flat spacetime by covariant derivatives $\nabla_{\mu}$ from curved spacetime \cite{Wald}. We will see later that when calculating the energy-momentum tensor symmetric of the spinorial field in the curved spacetime, it results in this same algebraic result of the equation (\ref{tensor_energia_momento}) replacing $\partial$ by $ \nabla$.

%%%%%%%%%%%%%%%%%%%%%%%%%%%%%%%%%%%%%%%%%%%%%%%%%%%%%%%%%%%%%%%%%%%%%%%%%%%%%%%%%%%%%%%%%%%%%%%%%%%%%%
%%%%%%%%%%%%%%%%%%%%%%%%%%%%%%%%%%%%%%%%%%%%%%%%%%%%%%%%%%%%%%%%%%%%%%%%%%%%%%%%%%%%%%%%%%%%%%%%%%%%%%%
%%%%%%%%%%%%%%%%%%%%%%%%%%%%%%%%%%%%%%%%%%%%%%%%%%%%%%%%%%%%%%%%%%%%%%%%%%%%%%%%%%%%%%%%%%%%%%%%%%%%%%
%%%%%%%%%%%%%%%%%%%%%%%%%%%%%%%%%%%%%%%%%%%%%%%%%%%%%%%%%%%%%%%%%%%%%%%%%%%%%%%%%%%%%%%%%%%%%%%%%%%%%%%
%%%%%%%%%%%%%%%%%%%%%%%%%%%%%%%%%%%%%%%%%%%%%%%%%%%%%%%%%%%%%%%%%%%%%%%%%%%%%%%%%%%%%%%%%%%%%%%%%%%%%%
%%%%%%%%%%%%%%%%%%%%%%%%%%%%%%%%%%%%%%%%%%%%%%%%%%%%%%%%%%%%%%%%%%%%%%%%%%%%%%%%%%%%%%%%%%%%%%%%%%%%%%%
%%%%%%%%%%%%%%%%%%%%%%%%%%%%%%%%%%%%%%%%%%%%%%%%%%%%%%%%%%%%%%%%%%%%%%%%%%%%%%%%%%%%%%%%%%%%%%%%%%%%%%
%%%%%%%%%%%%%%%%%%%%%%%%%%%%%%%%%%%%%%%%%%%%%%%%%%%%%%%%%%%%%%%%%%%%%%%%%%%%%%%%%%%%%%%%%%%%%%%%%%%%%%%
%%%%%%%%%%%%%%%%%%%%%%%%%%%%%%%%%%%%%%%%%%%%%%%%%%%%%%%%%%%%%%%%%%%%%%%%%%%%%%%%%%%%%%%%%%%%%%%%%%%%%%
%%%%%%%%%%%%%%%%%%%%%%%%%%%%%%%%%%%%%%%%%%%%%%%%%%%%%%%%%%%%%%%%%%%%%%%%%%%%%%%%%%%%%%%%%%%%%%%%%%%%%%%
%%%%%%%%%%%%%%%%%%%%%%%%%%%%%%%%%%%%%%%%%%%%%%%%%%%%%%%%%%%%%%%%%%%%%%%%%%%%%%%%%%%%%%%%%%%%%%%%%%%%%%
%%%%%%%%%%%%%%%%%%%%%%%%%%%%%%%%%%%%%%%%%%%%%%%%%%%%%%%%%%%%%%%%%%%%%%%%%%%%%%%%%%%%%%%%%%%%%%%%%%%%%%%

\section{Gauge Theory of Gravitation} \label{secao_Teoria_Gauge_Gravitacao}

%%%%%%%%%%%%%%%%%%%%%%%%%%%%%%%%%%%%%%%%%%%%%%%%%%%%%%%%%%%%%%%%%%%%%%%%%%%%%%%%%%%%%%%%%%%%%%%%%%%%%%
%%%%%%%%%%%%%%%%%%%%%%%%%%%%%%%%%%%%%%%%%%%%%%%%%%%%%%%%%%%%%%%%%%%%%%%%%%%%%%%%%%%%%%%%%%%%%%%%%%%%%%%
%%%%%%%%%%%%%%%%%%%%%%%%%%%%%%%%%%%%%%%%%%%%%%%%%%%%%%%%%%%%%%%%%%%%%%%%%%%%%%%%%%%%%%%%%%%%%%%%%%%%%%
%%%%%%%%%%%%%%%%%%%%%%%%%%%%%%%%%%%%%%%%%%%%%%%%%%%%%%%%%%%%%%%%%%%%%%%%%%%%%%%%%%%%%%%%%%%%%%%%%%%%%%%
%%%%%%%%%%%%%%%%%%%%%%%%%%%%%%%%%%%%%%%%%%%%%%%%%%%%%%%%%%%%%%%%%%%%%%%%%%%%%%%%%%%%%%%%%%%%%%%%%%%%%%
%%%%%%%%%%%%%%%%%%%%%%%%%%%%%%%%%%%%%%%%%%%%%%%%%%%%%%%%%%%%%%%%%%%%%%%%%%%%%%%%%%%%%%%%%%%%%%%%%%%%%%%
%%%%%%%%%%%%%%%%%%%%%%%%%%%%%%%%%%%%%%%%%%%%%%%%%%%%%%%%%%%%%%%%%%%%%%%%%%%%%%%%%%%%%%%%%%%%%%%%%%%%%%
%%%%%%%%%%%%%%%%%%%%%%%%%%%%%%%%%%%%%%%%%%%%%%%%%%%%%%%%%%%%%%%%%%%%%%%%%%%%%%%%%%%%%%%%%%%%%%%%%%%%%%%
%%%%%%%%%%%%%%%%%%%%%%%%%%%%%%%%%%%%%%%%%%%%%%%%%%%%%%%%%%%%%%%%%%%%%%%%%%%%%%%%%%%%%%%%%%%%%%%%%%%%%%
%%%%%%%%%%%%%%%%%%%%%%%%%%%%%%%%%%%%%%%%%%%%%%%%%%%%%%%%%%%%%%%%%%%%%%%%%%%%%%%%%%%%%%%%%%%%%%%%%%%%%%%
%%%%%%%%%%%%%%%%%%%%%%%%%%%%%%%%%%%%%%%%%%%%%%%%%%%%%%%%%%%%%%%%%%%%%%%%%%%%%%%%%%%%%%%%%%%%%%%%%%%%%%
%%%%%%%%%%%%%%%%%%%%%%%%%%%%%%%%%%%%%%%%%%%%%%%%%%%%%%%%%%%%%%%%%%%%%%%%%%%%%%%%%%%%%%%%%%%%%%%%%%%%%%%

In the year 1954, Yang and Mills \cite{Yang} introduced a non-Abelian group invariance $SU(2)$ to the spinorial action. At that time, non-abelian gauge theory was only a
mathematical theory, but today it is a central theory in Elementary Particle Physics, describing the electroweak force and strong nuclear interactions through internal groups that maintain the invariance of the action functional. Two years after the publication of Yang and Mills in 1956, Ryoyu Utiyama published a work about the gauge theory, even more comprehensive than the initial work of Yang and Mills, because Utiyama elaborates the theory of gauge for all semisimple Lie groups  and goes further in formulating gauge theory for gravitation, and subsequently Sciama and Kibble refined  through the invariance of the Poincar\'e extern group acting in the Minkowski spacetime \cite{Hehl_1,Utiyama, Kibble, Sciama, Hehl_2,Hehl_3}.

In a Yang-Mills theory we make a phase shift in the field to be invariant under some internal group such as $U(1)$, $SU(2)$ or $SU(3)$ 
for example, and we obtain the interaction of the fermionic field with the electromagnetism, isospin or quantum chromodynamics respectively. To obtain the interaction of the spinorial field with gravitation the gauge is made under the Lorentz transformations according to equation (\ref{transformacao_Lorentz}), 
\begin{equation}
\label{transformacao_espinor_1}
 \psi'(x')=\exp\left[\frac{i}{2}\epsilon^{\kappa\lambda}\Sigma_{\kappa\lambda}\right]\psi(x),
\end{equation}
where the transformation parameters are antisymmetric, $\epsilon_{\mu\nu}=-\epsilon_{\nu\mu}$ and $\Sigma_{\mu\nu}$ 
are the generators of the Lorentz group $SO(1,3)$. The gauge is initialized on this external group $SO(1,3)$.
We can rewrite the above transformation in the form
\begin{equation}
\label{transformacao_espinor_2}
 \psi'={\bm U}\psi,
\end{equation}
where
\begin{equation}
\label{transformacao_U_1}
{\bm U} = \exp\left[\frac{i}{2}\epsilon^{\kappa\lambda}\Sigma_{\kappa\lambda}\right].
\end{equation}
The Lorentz group transformation generators  $SO(3,1)$ 
are given in terms of the Dirac matrices, as seen in equation (\ref{gerador_Lorentz}),
\begin{equation}
 \Sigma_{\kappa\lambda}=\frac{i}{4}(\gamma_{\kappa}\gamma_{\lambda}-\gamma_{\lambda}\gamma_{\kappa})
 =\frac{i}{4}[\gamma_{\kappa},\gamma_{\lambda}],
\end{equation}
that obey the Lie algebra
\begin{equation}
\label{algebra_Lie}
 i[\Sigma_{\kappa\lambda}, \Sigma_{\mu\nu}] = g_{\mu\lambda}\Sigma_{\kappa\nu} -
  g_{\mu\kappa}\Sigma_{\lambda\nu} + g_{\nu\lambda}\Sigma_{\mu\kappa} -g_{\nu\kappa}\Sigma_{\mu\lambda}.
\end{equation}

It will be necessary to use the simplification of the expression below,
\begin{equation}
  i\epsilon^{\kappa\lambda}[\Sigma_{\kappa\lambda}, \Sigma_{\mu\nu}].
\end{equation}
as an exercise we can calculate it using equation (\ref{algebra_Lie}),
\begin{equation}
  i\epsilon^{\kappa\lambda}[\Sigma_{\kappa\lambda}, \Sigma_{\mu\nu}] =
  \epsilon^{\kappa\lambda}g_{\mu\lambda}\Sigma_{\kappa\nu} -
  \epsilon^{\kappa\lambda}g_{\mu\kappa}\Sigma_{\lambda\nu} +
 \epsilon^{\kappa\lambda} g_{\nu\lambda}\Sigma_{\mu\kappa} -
 \epsilon^{\kappa\lambda}g_{\nu\kappa}\Sigma_{\mu\lambda}
 = 
  {\epsilon^{\kappa}}_{\mu}\Sigma_{\kappa\nu} -
  {\epsilon_{\mu}}^{\lambda}\Sigma_{\lambda\nu} +
 {\epsilon^{\kappa}}_{\nu}\Sigma_{\mu\kappa} -
 {\epsilon_{\nu}}^{\lambda}\Sigma_{\mu\lambda},\nonumber
\end{equation}
knowing that  $\epsilon_{\mu\nu}=-\epsilon_{\nu\mu}$ 
so that 
$g^{\mu\kappa}\epsilon_{\mu\nu}=-g^{\mu\kappa}\epsilon_{\nu\mu}$ 
results in
${\epsilon^{\kappa}}_{\nu}=-{\epsilon_{\nu}}^{\kappa}$, 
so that
\begin{equation}
\label{algebra_Lie_2}
  i\epsilon^{\kappa\lambda}[\Sigma_{\kappa\lambda}, \Sigma_{\mu\nu}] = 
   {\epsilon^{\kappa}}_{\mu}\Sigma_{\kappa\nu} +
  {\epsilon^{\lambda}}_{\mu}\Sigma_{\lambda\nu} 
 -{\epsilon_{\nu}}^{\kappa}\Sigma_{\mu\kappa} 
 -{\epsilon_{\nu}}^{\lambda}\Sigma_{\mu\lambda}
 = 2 ({\epsilon^{\kappa}}_{\mu}\Sigma_{\kappa\nu} - {\epsilon_{\nu}}^{\kappa}\Sigma_{\mu\kappa}).
\end{equation}

The adjoint spinor $\bar{\psi} = \psi^{\dag}\gamma^0$ is transformed as
\begin{equation}
\label{transformacao_espinor_3}
 \bar{\psi'}=\bar{\psi}{\bm U}^{\dag},
\end{equation}
so that the term $\bar{\psi} \psi$ be invariant of Lorentz by the transformation 
(\ref{transformacao_espinor_1}),
\begin{equation}
\label{invariancia_espinor_1}
 \bar{\psi'}\psi'=(\bar{\psi}{\bm U}^{\dag}) ({\bm U}\psi) = \bar{\psi}\psi,
\end{equation}
where we must have
\begin{equation}
 {\bm U}^{\dag} {\bm U} =1,
\end{equation}
them ${\bm U}^{\dag} =  {\bm U}^{-1}$.

Now let us look at the Lagrangian of the spinorial or Dirac field
\begin{equation}
 {\cal L}= \frac{i}{2} \left(\bar\psi\gamma^{\mu}\partial_{\mu}\psi - 
 (\partial_{\mu}\bar\psi)\gamma^{\mu}\psi\right)
 + m\bar\psi \psi.
\end{equation}
Where the $\gamma^{\mu}$ matrices obey Clifford algebra in curved spacetime,
\begin{equation}
\label{Clifford_1}
 \{\gamma^{\mu},\gamma^{\nu}\}= \gamma^{\mu}\gamma^{\nu}+ \gamma^{\nu}\gamma^{\mu}=2g^{\mu\nu} 1\!\!1.
\end{equation}
being $1\!\!1$ a $4\times 4$ identity matrix.

So when we change the referential, the term $m\bar\psi \psi$ will be invariant according to the result
(\ref{invariancia_espinor_1}). But the kinetic terms will not be invariant by transformations of Lorentz. Let us look at the term $\left(\bar\psi\gamma^{\mu}\partial_{\mu}\psi\right)$,
\begin{equation}
 \bar\psi'\gamma^{\mu}\partial_{\mu}\psi' = 
  (\bar\psi {\bm U^{-1}})\gamma^{\mu}\partial_{\mu}({\bm U}\psi) 
 =    \bar\psi {\bm U^{-1}}\gamma^{\mu}[(\partial_{\mu}{\bm U})\psi+
{\bm U} \partial_{\mu}\psi]
= \bar\psi\gamma^{\mu}\partial_{\mu}\psi  +  
 \bar\psi\gamma^{\mu} {\bm U^{-1}}(\partial_{\mu}{\bm U})\psi,
\end{equation}
where it is clear that the second term breaks the invariance of the Lagrangian. Then it is necessary to obtain a covariant derivative,
\begin{equation}
\label{del_covariante_espinor_1}
\bm D'_{\mu}\psi'= {\bm U}\bm D_{\mu}\psi, 
\end{equation}
so that the kinetic term with covariant derivative,
$ \bar\psi\gamma^{\mu}\bm D_{\mu}\psi $, under a Lorentz transformation results in
\begin{equation}
 \bar\psi'\gamma^{\mu}\bm D'_{\mu}\psi' =  (\bar\psi {\bm U^{-1}})({\bm U}\bm D_{\mu}\psi)= 
 \bar\psi\gamma^{\mu}\bm D_{\mu}\psi.
\end{equation}
In this way the kinetic term will be invariant by the transformation (\ref{transformacao_espinor_1}). 
Note also that of expression (\ref{del_covariante_espinor_1}) results in
\begin{equation}
\bm D'_{\mu}\psi'= {\bm U}\bm D_{\mu}({\bm{U}}^{-1}\psi'),
\end{equation}
where we use the expression (\ref{transformacao_espinor_2}), with $\bm{U}^{-1}\psi'=\psi$,  so that the transformation of the covariant derivative of a coordinate system ${\cal O}$ to another coordinate system ${\cal O}'$ is given by
\begin{equation}
 \label{del_covariante_espinor_2}
 \bm D'_{\mu} = {\bm U}\bm D_{\mu}{\bm{U}}^{-1}.
\end{equation}

To obtain the interaction of the spinorial field with gravity we express the Lagrangian in terms of a tetrad or vierbein on a non-coordinate basis given by
\begin{equation}
%\label{base_NC1}
{\tilde{\bm e}}_{\alpha}={e_{\alpha}}^{\mu}\bm E_{\mu}
\end{equation} 
with the respective differential forms given by
\begin{equation}
%\label{base_NC2}
 \tilde{\bm\theta}^{\beta}={\omega^{\beta}}_{\nu}dx^{\nu}.
\end{equation}
where $({e_{\alpha}}^{\mu})$ and $({\omega^{\beta}}_{\nu})\in GL(4,\mathbb{R})$ with 
\begin{equation}
{\omega^{\beta}}_{\nu}{e_{\alpha}}^{\nu} = {\delta^{\beta}}_{\alpha} \hspace*{1cm} \mbox{and} \hspace*{1cm} {\omega^{\beta}}_{\nu}{e_{\beta}}^{\mu} = {\delta_{\nu}}_{\mu}.
\end{equation}
For a more detailed review on non-coordinated basis see reference \cite{wytler}
In terms of a non-coordinate orthonormal basis, the Dirac matrices are given by,
\begin{equation}
 \gamma^{\mu} = \gamma^{\alpha} {e_{\alpha}}^{\mu} \hspace*{1cm} \mbox{and} \hspace*{1cm}  \gamma_{\alpha}= {e_{\alpha}}^{\mu}\gamma_{\mu},
\end{equation}
so that the Clifford algebra for the $\gamma_{\alpha}$ matrices are 
\begin{equation}
\{  \gamma_{\alpha}, \gamma_{\beta}\} =  \gamma_{\alpha} \gamma_{\beta} + \gamma_{\beta}\gamma_{\alpha} = 
  {e_{\alpha}}^{\mu}\gamma^{\mu}{e_{\beta}}^{\nu}\gamma^{\nu} + {e_{\beta}}^{\nu}\gamma^{\nu} 
 {e_{\alpha}}^{\mu}\gamma^{\mu} = \{  \gamma_{\mu}, \gamma_{\nu}\}{e_{\alpha}}^{\mu} {e_{\beta}}^{\nu},
 \nonumber
\end{equation}
where the Clifford  algebra (\ref{Clifford_1}) can be used leading to
\begin{equation}
   \{  \gamma_{\alpha}, \gamma_{\beta}\} = 2g_{\mu\nu}{e_{\alpha}}^{\mu} {e_{\beta}}^{\nu}1\!\!1,\nonumber
\end{equation}
or else
\begin{equation}
   \{  \gamma_{\alpha}, \gamma_{\beta}\} = 2\eta_{\alpha\beta} 1\!\!1,  \nonumber
\end{equation}
where we must identify that 
\begin{equation}
\label{vierbeins}
 g_{\mu\nu}{e_{\alpha}}^{\mu} {e_{\beta}}^{\nu} = \eta_{\alpha\beta}.
\end{equation}

The covariant derivative will be given by
\begin{equation}
\bm D_{\mu}= {\omega^{\beta}}_{\mu}\bm D_{\beta},   
\end{equation}
so that the kinetic term of the Lagrangian of Dirac $\frac{i}{2} \bar\psi \gamma^{\mu}\bm D_{\mu}\psi$ is given by
\begin{equation}
\frac{i}{2} \bar\psi \gamma^{\mu}\bm D_{\mu}\psi =  
\frac{i}{2} \bar\psi \gamma^{\alpha}{e_{\alpha}}^{\mu}  {\omega^{\beta}}_{\mu}\bm D_{\beta}\psi =
\frac{i}{2} \bar\psi \gamma^{\alpha}{\delta_{\alpha}}^{\beta}\bm D_{\beta}\psi =
\frac{i}{2} \bar\psi \gamma^{\alpha}\bm D_{\alpha}\psi.
\end{equation}

From equation (\ref{del_covariante_espinor_2}), where $ \bm D'_{\mu} = {\bm U}\bm D_{\mu}{\bm{U}}^{-1} $, we have that
\begin{eqnarray}
 \bm D'_{\mu} ={\omega'^{\beta}}_{\mu}\bm D'_{\beta} = {\bm U}\bm D_{\mu}{\bm{U}}^{-1}, \nonumber
\end{eqnarray}
that multiplying by ${e'_{\alpha}}^{\mu}$ results in
\begin{equation}
 {e'_{\alpha}}^{\mu}{\omega'^{\beta}}_{\mu}\bm D'_{\beta} = {e'_{\alpha}}^{\mu}{\bm U}\bm D_{\mu}{\bm{U}}^{-1},\nonumber
\end{equation}
knowing that the Lorentz transformation for  ${e'_{\alpha}}^{\mu}$ is given by
${e'_{\alpha}}^{\mu} = {\Lambda_{\alpha}}^{\beta}{e_{\beta}}^{\mu}$ \cite{wytler}, we have
\begin{eqnarray}
 {\delta_{\alpha}}^{\beta}\bm D'_{\beta} &=&  
 {\Lambda_{\alpha}}^{\beta}{e_{\beta}}^{\mu}{\bm U}\bm D_{\mu}{\bm{U}}^{-1}\cr
 \bm D'_{\alpha} &=&  {\Lambda_{\alpha}}^{\beta}{\bm U}{e_{\beta}}^{\mu}\bm D_{\mu}{\bm{U}}^{-1}   ,\nonumber
\end{eqnarray}
so that we have the result,
\begin{equation}
\label{del_covariante_espinor_3}
 \bm D'_{\alpha} =  {\Lambda_{\alpha}}^{\beta}{\bm U}\bm D_{\beta}{\bm{U}}^{-1}   .
\end{equation}

Now let the covariant derivative be given by
\begin{equation}
 \label{del_covariante_espinor_4}
 \bm D_{\alpha} = {e_{\alpha}}^{\mu}(\partial_{\mu}+\Omega_{\mu}), 
\end{equation}
then let us see how this covariant derivative transforms under the Lorentz transformations given by the expression
(\ref{del_covariante_espinor_3}), such that
\begin{eqnarray}
 \bm D'_{\alpha} \psi &=&  
 {\Lambda_{\alpha}}^{\beta}{\bm U}{e_{\beta}}^{\mu}(\partial_{\mu}+\Omega_{\mu}){\bm{U}}^{-1}\psi  \cr
 &=&  {\Lambda_{\alpha}}^{\beta}{\bm U}{e_{\beta}}^{\mu}[(\partial_{\mu}{\bm{U}}^{-1}) \psi
 +{\bm{U}}^{-1}\partial_{\mu}\psi+ \Omega_{\mu}{\bm{U}}^{-1}\psi]\cr
 &=& {\Lambda_{\alpha}}^{\beta}{\bm U}{e_{\beta}}^{\mu}[-{\bm{U}}^{-2}(\partial_{\mu}{\bm{U}}) \psi
 +{\bm{U}}^{-1}\partial_{\mu}\psi+ \Omega_{\mu}{\bm{U}}^{-1} \psi]\cr
  &=& {\Lambda_{\alpha}}^{\beta}{e_{\beta}}^{\mu}[{\bm U}{\bm{U}}^{-1}\partial_{\mu}\psi
 + {\bm U}\Omega_{\mu}{\bm{U}}^{-1}\psi
  -  {\bm U}{\bm{U}}^{-2}(\partial_{\mu}{\bm{U}}) \psi]\cr
  &=&{e'_{\alpha}}^{\mu}(\partial_{\mu}
  +  {\bm U}\Omega_{\mu}{\bm{U}}^{-1}
  -  {\bm{U}}^{-1}\partial_{\mu}{\bm{U}}) \psi,
\end{eqnarray}
and making the identification
\begin{equation}
 \label{del_covariante_espinor_4a}
  \bm D'_{\alpha}  ={e'_{\alpha}}^{\mu}(\partial_{\mu}+\Omega'_{\mu}),
\end{equation}
thus, we have that
\begin{equation}
\label{Omega'}
 \Omega'_{\mu} = {\bm U}\Omega_{\mu}{\bm{U}}^{-1}
  -  {\bm{U}}^{-1}\partial_{\mu}{\bm{U}}.
\end{equation}

It is important and straightforward to compare the expression (\ref{del_covariante_espinor_4}) with (\ref{del_covariante_espinor_4a}) so that,
\begin{equation}
 \partial_{\alpha} = {e_{\alpha}}^{\mu}\partial_{\mu} \nonumber
\end{equation}
and additionally
\begin{equation}
 \partial\, '_{\alpha} = {e'_{\alpha}}^{\mu}\partial_{\mu}, \nonumber
\end{equation}
so that the expression (\ref{del_covariante_espinor_4}) can be written as
\begin{eqnarray}
 \bm D _{\alpha} = {e_{\alpha}}^{\mu}(\partial_{\mu}+\Omega_{\mu})=
 \partial_{\alpha} +{e_{\alpha}}^{\mu}\Omega_{\mu},\nonumber
\end{eqnarray}
which in the coordinate system ${\cal O}'$ is given by
\begin{eqnarray}
 \bm D\,' _{\alpha} = \partial\,'_{\alpha} +{e'_{\alpha}}^{\mu}\Omega\,'_{\mu}= {e'_{\alpha}}^{\mu}\partial_{\mu}  +{e'_{\alpha}}^{\mu}\Omega\,'_{\mu} = {e'_{\alpha}}^{\mu}(\partial_{\mu}+ \Omega\,'_{\mu}),\nonumber
\end{eqnarray}
which is in agreement with the result (\ref{del_covariante_espinor_4a}).

From the Lorentz transformation to the spinor given by the expression (\ref{transformacao_U_1}) where the term 
$\epsilon^{\alpha\beta}$ is the infinitesimal parameters of transformations,
 we have that
\begin{equation}
\label{transformacao_U_2}
{\bm U} = \exp\left[\frac{i}{2}\epsilon^{\alpha\beta}\Sigma_{\alpha\beta}\right]
\approx 1 +\frac{i}{2}\epsilon^{\alpha\beta}\Sigma_{\alpha\beta},
\end{equation}
So replacing this infinitesimal transformation in the expression (\ref{Omega'}) and neglecting second-order terms in $\epsilon$, we have
\begin{eqnarray}
\label{Omega'1}
  \Omega'_{\mu} &=&\left(1 +\frac{i}{2}\epsilon^{\alpha\beta}\Sigma_{\alpha\beta}\right)\Omega_{\mu}
  \left(1 -\frac{i}{2}\epsilon^{\gamma\delta}\Sigma_{\gamma\delta}\right)
  - \left(1 -\frac{i}{2}\epsilon^{\gamma\delta}\Sigma_{\gamma\delta}\right)\partial_{\mu}
  \left(1 +\frac{i}{2}\epsilon^{\alpha\beta}\Sigma_{\alpha\beta}\right)\cr
  &=&\left(1 +\frac{i}{2}\epsilon^{\alpha\beta}\Sigma_{\alpha\beta}\right)
    \left(\Omega_{\mu} -\frac{i}{2}\Omega_{\mu}\epsilon^{\gamma\delta}\Sigma_{\gamma\delta}\right)
    - \left(1 -\frac{i}{2}\epsilon^{\gamma\delta}\Sigma_{\gamma\delta}\right)\partial_{\mu}
    \frac{i}{2}\epsilon^{\alpha\beta}\Sigma_{\alpha\beta}\cr
 &=&    \Omega_{\mu}  + \frac{i}{2}\epsilon^{\alpha\beta}\Sigma_{\alpha\beta}\Omega_{\mu}
 - \frac{i}{2}\epsilon^{\gamma\delta}\Omega_{\mu}\Sigma_{\gamma\delta} - \partial_{\mu}
    \frac{i}{2}\epsilon^{\alpha\beta}\Sigma_{\alpha\beta}\cr
   &=&    \Omega_{\mu}  + \frac{i}{2}\epsilon^{\alpha\beta}\left(\Sigma_{\alpha\beta}\Omega_{\mu}  
   -\Omega_{\mu} \Sigma_{\alpha\beta}\right)-\frac{i}{2} (\partial_{\mu}\epsilon^{\alpha\beta})\Sigma_{\alpha\beta}\cr
    &=&    \Omega_{\mu}  + \frac{i}{2}\epsilon^{\alpha\beta}[ \Sigma_{\alpha\beta},\Omega_{\mu}] 
    -\frac{i}{2} (\partial_{\mu}\epsilon^{\alpha\beta})\Sigma_{\alpha\beta}.
\end{eqnarray}
It is possible to identify the geometric element that transforms under Lorentz transformation like $\Omega_{\mu}$ in the above equation. In the references \cite{Nakahara, wytler} we have the Lorentz transformations of the one-form connection,
\begin{equation}
\label{conexão_NC1.0}
 {{\bm\Gamma}'^{\alpha}}_{\beta} = 
 {\Lambda^{\alpha}}_{\gamma}  {{\bm\Gamma}^{\gamma}}_{\delta} {(\Lambda^{-1})^{\delta}}_{\beta}
+ {\Lambda^{\alpha}}_{\gamma}d({\Lambda^{-1})^{\gamma}}_{\beta}.
\end{equation}
Using a local coordinate infinitesimal transformation, where 
$$ {\Lambda^{\alpha}}_{\gamma} \approx  {\delta^{\alpha}}_{\gamma}  +  {\epsilon^{\alpha}}_{\gamma}, $$
we have the infinitesimal Lorentz transformations for the one-form connection
\begin{eqnarray}
\label{conexão_NC1.1}
 {{\bm\Gamma}'^{\alpha}}_{\beta} &=& 
 ({\delta^{\alpha}}_{\gamma}  +  {\epsilon^{\alpha}}_{\gamma})  {{\bm\Gamma}^{\gamma}}_{\delta} 
({\delta^{\delta}}_{\beta}  -  {\epsilon^{\delta}}_{\beta})
+ ({\delta^{\alpha}}_{\gamma}  +  {\epsilon^{\alpha}}_{\gamma}) 
d({\delta^{\gamma}}_{\beta}  -  {\epsilon^{\gamma}}_{\beta}) \cr
&=&  ({\delta^{\alpha}}_{\gamma}  +  {\epsilon^{\alpha}}_{\gamma})( {{\bm\Gamma}^{\gamma}}_{\beta} 
 - {{\bm\Gamma}^{\gamma}}_{\delta}   {\epsilon^{\delta}}_{\beta}) + 
 ({\delta^{\alpha}}_{\gamma}  -  {\epsilon^{\alpha}}_{\gamma})d  {\epsilon^{\gamma}}_{\beta}\cr
 &=&  {{\bm\Gamma}^{\alpha}}_{\beta} + {\epsilon^{\alpha}}_{\gamma}{{\bm\Gamma}^{\gamma}}_{\beta} 
 - {{\bm\Gamma}^{\alpha}}_{\delta}   {\epsilon^{\delta}}_{\beta} -d  {\epsilon^{\alpha}}_{\beta}.
\end{eqnarray}
From the definition of one-form we have that  
\begin{equation}
\label{conexão_NC1.2}
 {\bm\Gamma^{\alpha}}_{\beta} \equiv  {\Gamma^{\alpha}}_{\gamma\beta}\tilde{\bm\theta}^{\gamma} =
  {\Gamma^{\alpha}}_{\gamma\beta}{\omega^{\gamma}}_{\mu}\,dx^{\mu}= {\Gamma^{\alpha}}_{\mu\beta}dx^{\mu},
\end{equation}
and replacing these values in the expression (\ref{conexão_NC1.1}) we obtain
\begin{equation}
 {\Gamma'^{\alpha}}_{\mu\beta} dx^{\mu}= {\Gamma^{\alpha}}_{\mu\beta} dx^{\mu} + {\epsilon^{\alpha}}_{\gamma}
 {\Gamma^{\gamma}}_{\mu\beta} dx^{\mu} - {\Gamma^{\alpha}}_{\mu\gamma} dx^{\mu} {\epsilon^{\gamma}}_{\beta}
 -\partial_{\mu}{\epsilon^{\alpha}}_{\beta} dx^{\mu}, \nonumber
\end{equation}
or else
\begin{equation}
 {\Gamma'^{\alpha}}_{\mu\beta} = {\Gamma^{\alpha}}_{\mu\beta}  + {\epsilon^{\alpha}}_{\gamma}
 {\Gamma^{\gamma}}_{\mu\beta} - {\Gamma^{\alpha}}_{\mu\gamma}  {\epsilon^{\gamma}}_{\beta}
 -\partial_{\mu}{\epsilon^{\alpha}}_{\beta} .
\end{equation}
Let us raise the index $\beta$ and multiply the above expression by the Lorentz transformation generator of spinors $\Sigma_{\alpha\beta}$,
\begin{eqnarray}
\label{Gauge_Grav_Espinor2}
 {{\Gamma'\,^{\alpha}}_{\mu}}^{\beta}\Sigma_{\alpha\beta} &=& {{\Gamma^{\alpha}}_{\mu}}^{\beta}\Sigma_{\alpha\beta}
 + {\epsilon^{\alpha}}_{\gamma} {{\Gamma^{\gamma}}_{\mu}}^{\beta}\Sigma_{\alpha\beta}
 - {\epsilon^{\gamma\beta}} {\Gamma^{\alpha}}_{\mu\gamma}  \Sigma_{\alpha\beta}
 -(\partial_{\mu}{\epsilon^{\alpha\beta}})\Sigma_{\alpha\beta}\cr
 &=&
 {{\Gamma^{\alpha}}_{\mu}}^{\beta}\Sigma_{\alpha\beta}
 + \underbrace{   {\epsilon^{\alpha}}_{\gamma}  {{\Gamma^{\gamma}}_{\mu}}^{\beta}\Sigma_{\alpha\beta}
 - {\epsilon_{\gamma}}^{\beta} {{\Gamma^{\alpha}}_{\mu}}^{\gamma}  \Sigma_{\alpha\beta}   }
 -(\partial_{\mu}{\epsilon^{\alpha\beta}})\Sigma_{\alpha\beta}.
\end{eqnarray}
It is possible to obtain an expression that will facilitate the identification of the covariant derivative,
for reviewing the terms underbraced above.

As an exercise we call the result of the equation (\ref{algebra_Lie_2})
\begin{equation}
 i\epsilon^{\alpha\beta}[\Sigma_{\alpha\beta}, \Sigma_{\gamma\delta}]= 
 2 ({\epsilon^{\alpha}}_{\gamma}\Sigma_{\alpha\delta} - {\epsilon_{\delta}}^{\alpha}\Sigma_{\gamma\alpha}),\nonumber
\end{equation}
and multiply by ${{\Gamma^{\gamma}}_{\mu}}^{\delta}$,
resulting in
\begin{eqnarray}
 i\epsilon^{\alpha\beta}[\Sigma_{\alpha\beta}, \Sigma_{\gamma\delta}]{{\Gamma^{\gamma}}_{\mu}}^{\delta} &=& 
 2 ({\epsilon^{\alpha}}_{\gamma}\Sigma_{\alpha\delta}{{\Gamma^{\gamma}}_{\mu}}^{\delta}
 - {\epsilon_{\delta}}^{\alpha}\Sigma_{\gamma\alpha}{{\Gamma^{\gamma}}_{\mu}}^{\delta})\cr
 \epsilon^{\alpha\beta}[\Sigma_{\alpha\beta},\frac{i}{2}{{\Gamma^{\gamma}}_{\mu}}^{\delta}\Sigma_{\gamma\delta}] &=&
 {\epsilon^{\alpha}}_{\gamma}{{\Gamma^{\gamma}}_{\mu}}^{\delta}\Sigma_{\alpha\delta}
 - {\epsilon_{\delta}}^{\alpha}{{\Gamma^{\gamma}}_{\mu}}^{\delta}\Sigma_{\gamma\alpha},\nonumber
\end{eqnarray}
changing some dummy indices, the above equation can be written as
\begin{equation}
 {\epsilon^{\alpha}}_{\gamma}{{\Gamma^{\gamma}}_{\mu}}^{\beta}\Sigma_{\alpha\beta}
 - {\epsilon_{\gamma}}^{\beta}{{\Gamma^{\alpha}}_{\mu}}^{\gamma}\Sigma_{\alpha\beta}=
 \epsilon^{\alpha\beta}[\Sigma_{\alpha\beta},\frac{i}{2}{{\Gamma^{\gamma}}_{\mu}}^{\delta}\Sigma_{\gamma\delta}].
\end{equation}
Then we can substitute the above result into equation (\ref{Gauge_Grav_Espinor2}),
\begin{equation}
 \label{Gauge_Grav_Espinor3}
 {{\Gamma'^{\alpha}}_{\mu}}^{\beta}\Sigma_{\alpha\beta} = {{\Gamma^{\alpha}}_{\mu}}^{\beta}\Sigma_{\alpha\beta}
 + \epsilon^{\alpha\beta}[\Sigma_{\alpha\beta},\frac{i}{2}{{\Gamma^{\gamma}}_{\mu}}^{\delta} \Sigma_{\gamma\delta}] 
 -(\partial_{\mu}{\epsilon^{\alpha\beta}})\Sigma_{\alpha\beta}.
\end{equation}
Now let us multiply this result by $\frac{i}{2}$ and compare it with the expression of $\Omega_{\mu}$ obtained in (\ref{Omega'1}),
\begin{equation}
 \begin{cases}
 \frac{i}{2} {{\Gamma'^{\alpha}}_{\mu}}^{\beta}\Sigma_{\alpha\beta} = 
 \frac{i}{2}{{\Gamma^{\alpha}}_{\mu}}^{\beta}\Sigma_{\alpha\beta}
 +\frac{i}{2} \epsilon^{\alpha\beta}[\Sigma_{\alpha\beta},\frac{i}{2}{{\Gamma^{\gamma}}_{\mu}}^{\delta}
 \Sigma_{\gamma\delta}] 
 -\frac{i}{2}(\partial_{\mu}{\epsilon^{\alpha\beta}})\Sigma_{\alpha\beta}\cr\cr
  \Omega'_{\mu} =
  \Omega_{\mu}  + \frac{i}{2}\epsilon^{\alpha\beta}[ \Sigma_{\alpha\beta},\Omega_{\mu}] 
    -\frac{i}{2} (\partial_{\mu}\epsilon^{\alpha\beta})\Sigma_{\alpha\beta}
 \end{cases}\nonumber
\end{equation}
where it is possible to identify
\begin{equation}
 \label{Omega=GammaSigma}
  \bm\Omega_{\mu} =  \frac{i}{2}{{\Gamma^{\alpha}}_{\mu}}^{\beta}\Sigma_{\alpha\beta}.
\end{equation}
Then the covariant derivative proposed in the expression of equation (\ref{del_covariante_espinor_4}) where
\begin{equation}
 \bm D_{\alpha} = {e_{\alpha}}^{\mu}(\partial_{\mu}+\bm\Omega_{\mu}) \nonumber
 \end{equation}
must be the following covariant derivative,
\begin{equation}
\label{del_covariante_espinor_5} 
 \bm D_{\alpha} = {e_{\alpha}}^{\mu}\left(\partial_{\mu}+ 
 \frac{i}{2}{{\Gamma^{\beta}}_{\mu}}^{\gamma}\Sigma_{\beta\gamma}\right), 
 \end{equation}
%%%%%%%%%%%%%%%%%%%%%%%%%%%%%%%%%%%%%%%%%%%%%%%%%%%%%%%%%%%%%%%%%%%%%%%%%%%%%%%%%%%%%%%%%%%%% 
or 
\begin{equation}
\label{Operador_curvatura_2}
\bm D_{\alpha} = \partial_{\alpha}+ \frac{i}{2}{{\Gamma^{\beta}}_{\alpha}}^{\gamma}\Sigma_{\beta\gamma}.
\end{equation}

In the structure of the Poincar\'e gauge field theory, denoted by $PG$, the matter spinorial field $\psi(x^{\mu})$ to be translated from $x^{\mu}$ to $x^{\mu} + \epsilon^{\mu}$ (where $\epsilon^{\mu}$ are  4 translation infinitesimal parameters) 
and with the fixed orientation, the translation generator for the spinorial field is the covariant derivative obtained in equation (\ref{del_covariante_espinor_5}), this being a parallel transport operation \cite{Hehl_3}. The covariant derivative (\ref{del_covariante_espinor_5}) 
is a translational type transformation and distinguishes Poincar\'e gauge field theory from Yang-Mills gauge theories which are transformations of internal symmetries when the field $\psi(x^{\mu})$ is moved to a different point in spacetime.

The above covariant derivative, when operating on a spinor  $\psi$, will have as generator of transformation of Lorentz the matrices
$\Sigma_{\alpha\beta}
 =\frac{i}{4}[\gamma_{\alpha},\gamma_{\beta}] $.
But this same covariant derivative can operate in other fields where we will have,
 \begin{itemize}
  \item  $\Sigma_{\alpha\beta} = 0$ 
  when the covariant derivative  $\bm D_{\alpha}$ 
is applied to a scalar field $\phi$; 
  \item $\Sigma_{\alpha\beta} \rightarrow [\Sigma_{\alpha\beta}]^{\delta}_{\gamma} = 
  i({\delta_{\beta}}^{\delta}\eta_{\alpha\gamma} - {\delta_{\alpha}}^{\delta}\eta_{\beta\gamma})$
  when $\bm D_{\alpha}$ is applied to a contravariant vector field $A^{\alpha}$ 
so that
  $A'^{\alpha}=A^{\alpha}+{\epsilon^{\alpha}}_{\beta}A^{\beta}$ 
for infinitesimal transformations.
While we will have 
 $\Sigma_{\alpha\beta} \rightarrow [\Sigma_{\alpha\beta}]^{\delta}_{\gamma} = 
  i( \eta_{\beta\gamma}{\delta_{\alpha}}^{\delta} - \eta_{\alpha\gamma}{\delta_{\beta}}^{\delta} )$
  when $\bm D_{\alpha}$ is applied to a covariant vector field $A_{\alpha}$ so that
$A'_{\alpha}=A_{\alpha}+{\epsilon_{\alpha}}^{\beta}A_{\beta}$ for infinitesimal transformations
 \end{itemize}
%%%%%%%%%%%%%%%%%%%%%%%%%%%%%%%%%%%%%%%%%%%%%%%%%%%%%%%%%%%%%%%%%%%%%%%%%%%%%%%%%%%%%%%%%%%%%%%%
%%%%%%%%%%%%%%%%%%%%%%%%%%%%%%%%%%%%%%%%%%%%%%%%%%%%%%%%%%%%%%%%%%%%%%%%%%%%%%%%%%%%%%%%%%%%%%%%
As an exercise we will calculate the covariant derivative of a vector in this orthonormal non-coordinate system,
\begin{equation}
 \bm D_{\alpha} V^{\beta} = \partial_{\alpha} V^{\beta} 
 + \frac{i}{2}{{\Gamma^{\gamma}}_{\alpha}}^{\delta}\Sigma_{\gamma\delta}V^{\beta},
\end{equation}
where $\Sigma_{\gamma\delta}V^{\beta}=[\Sigma_{\gamma\delta}]^{\beta}_{\epsilon}V^{\epsilon} $, 
observing that the vector is contravariant, so that,
\begin{equation}
 \bm D_{\alpha} V^{\beta} =  \partial_{\alpha} V^{\beta} +
\frac{i}{2}{{\Gamma^{\gamma}}_{\alpha}}^{\delta} \,\,
i({\delta_{\delta}}^{\beta}\eta_{\gamma\epsilon} - {\delta_{\gamma}}^{\beta}\eta_{\delta\epsilon}) V^{\epsilon} 
= \partial_{\alpha} V^{\beta} - \frac{1}{2}{{\Gamma^{}}_{\epsilon\alpha}}^{\beta} V^{\epsilon} 
+ \frac{1}{2} {{\Gamma^{\beta}}_{\alpha\epsilon}} V^{\epsilon}, \nonumber
\end{equation}
in the orthonormal non-coordinate system we have that
 ${{\Gamma^{}}_{\epsilon\alpha}}^{\beta} = -{{\Gamma^{\beta}}_{\alpha\epsilon}}  $ \cite{wytler} that results in
\begin{equation}
  \bm D_{\alpha} V^{\beta} = \partial_{\alpha} V^{\beta}+V^{\gamma}{{\Gamma^{\beta}}_{\alpha\gamma}}. 
\end{equation}
We now observe that the covariant derivative $\bm D_{\alpha}$ performs the same
covariant derivative in a manifold with a metric $g_{\alpha\beta}$,
$$ \nabla_{\alpha} V^{\beta} = \partial_{\alpha} V^{\beta}+V^{\gamma}{{\Gamma^{\beta}}_{\alpha\gamma}}.$$
From this point on we will change the symbol of the covariant derivative of gauge by the affin connection $\nabla$. In the appendix the details of how to obtain the curvature term from the covariant derivative are shown.

%%%%%%%%%%%%%%%%%%%%%%%%%%%%%%%%%%%%%%%%%%%%%%%%%%%%%%%%%%%%%%%%%%%%%%%%%%%%%%%%%%%%%%%%%%%%%%%% 
 The action for a spinor field in the curved spacetime must be given by
 \begin{equation}
 S_{M}= \int_{}d^4x \sqrt{-g}\left\{\frac{i}{2} \left[\bar\psi\gamma^{\mu}\nabla_{\mu}\psi - 
 (\nabla_{\mu}\bar{\psi})\gamma^{\mu}\psi\right]
 +m\bar\psi \psi\right\},
\end{equation} 
where we must draw attention to the term $\nabla_{\mu}\bar{\psi} = \partial_{\mu}\bar{\psi} - 
 \frac{i}{2}{{\Gamma^{\gamma}}_{\mu}}^{\delta}\Sigma_{\gamma\delta}\bar{\psi}$
and using the identity $\sqrt{-g}=\det({e_{\alpha}}^{\mu})=\det(e)$, 
we obtain the invariant action for spinor field under general changes of coordinates 
\begin{equation}
\label{acao_Dirac_gravitacao_2}
 S_{M}= \int_{\Omega}d^4 x\,\,\det(e) \left\{\frac{i}{2} \left[\bar\psi\gamma^{\alpha}\,\,{e_{\alpha}}^{\mu}
\left(\partial_{\mu}+ 
 \frac{i}{2}{{\Gamma^{\gamma}}_{\mu}}^{\delta}\Sigma_{\gamma\delta}\right)\psi -  {e_{\alpha}}^{\mu}
 \left(\partial_{\mu}\bar\psi- 
 \frac{i}{2}{{\Gamma^{\gamma}}_{\mu}}^{\delta}\Sigma_{\gamma\delta}\bar\psi\right)\gamma^{\alpha}\psi\right]
 +m\bar\psi \psi\right\}.
\end{equation}

%%%%%%%%%%%%%%%%%%%%%%%%%%%%%%%%%%%%%%%%%%%%%%%%%%%%%%%%%%%%%%%%%%%%%%%%%%%%%%%%%%%%%

%%%%%%%%%%%%%%%%%%%%%%%%%%%%%%%%%%%%%%%%%%%%%%%%%%%%%%%%%%%%%%%%%%%%%%%%%%%%%%%%%%%%%

%%%%%%%%%%%%%%%%%%%%%%%%%%%%%%%%%%%%%%%%%%%%%%%%%%%%%%%%%%%%%%%%%%%%%%%%%%%%%%%%%%%%%

%%%%%%%%%%%%%%%%%%%%%%%%%%%%%%%%%%%%%%%%%%%%%%%%%%%%%%%%%%%%%%%%%%%%%%%%%%%%%%%%%%%%%

%%%%%%%%%%%%%%%%%%%%%%%%%%%%%%%%%%%%%%%%%%%%%%%%%%%%%%%%%%%%%%%%%%%%%%%%%%%%%%%%%%%%%

%%%%%%%%%%%%%%%%%%%%%%%%%%%%%%%%%%%%%%%%%%%%%%%%%%%%%%%%%%%%%%%%%%%%%%%%%%%%%%%%%%%%%

%%%%%%%%%%%%%%%%%%%%%%%%%%%%%%%%%%%%%%%%%%%%%%%%%%%%%%%%%%%%%%%%%%%%%%%%%%%%%%%%%%%%%

%%%%%%%%%%%%%%%%%%%%%%%%%%%%%%%%%%%%%%%%%%%%%%%%%%%%%%%%%%%%%%%%%%%%%%%%%%%%%%%%%%%%%

%%%%%%%%%%%%%%%%%%%%%%%%%%%%%%%%%%%%%%%%%%%%%%%%%%%%%%%%%%%%%%%%%%%%%%%%%%%%%%%%%%%%%

%%%%%%%%%%%%%%%%%%%%%%%%%%%%%%%%%%%%%%%%%%%%%%%%%%%%%%%%%%%%%%%%%%%%%%%%%%%%%%%%%%%%%

%%%%%%%%%%%%%%%%%%%%%%%%%%%%%%%%%%%%%%%%%%%%%%%%%%%%%%%%%%%
%%%%%%%%%%%%%%%%%%%%%%%%%%%%%%%%%%%%%%%%%%%%%%%%%%%%%%%%%%%
%%%%%%%%%%%%%%%%%%%%%%%%%%%%%%%%%%%%%%%%%%%%%%%%%%%%%%%%%%%

\section{Energy-momentum and spin current density tensors for the spinorial field} \label{secao_tensor_energia-momento}

We have seen that the energy-momentum tensor due to the presence of a field is given by the expression 
(\ref{Tensor_E-P_1}), 
\begin{equation}
T^{\mu\nu} = \frac{2}{\sqrt{-g}}\frac{\delta S_{M} }{\delta g_{\mu\nu}}.\nonumber
\end{equation}
This expression is useful to find the energy-momentum tensor when the $S_{M}$ action of the field is given in terms of metric tensor $g_{\mu\nu}$, as is the case of the scalar field and the vector field. But in the case of the spinorial field
the action is given in terms of the vierbein ${e_{\alpha}}^{\mu}$ as seen in the equation (\ref{acao_Dirac_gravitacao_2}),
%\begin{footnotesize}
\begin{equation}
 S_{M}= \int_{\Omega}d^4 x\,\,\det(e) \left\{\frac{i}{2} \left[\bar\psi\gamma^{\alpha}\,\,{e_{\alpha}}^{\mu}
\left(\partial_{\mu}+ 
 \frac{i}{2}{\Gamma^{\gamma\delta}}_{\mu}\Sigma_{\gamma\delta}\right)\psi -  {e_{\alpha}}^{\mu}
 \left(\partial_{\mu}\bar\psi- 
 \frac{i}{2}{\Gamma^{\gamma\delta}}_{\mu}\Sigma_{\gamma\delta}\bar\psi\right)\gamma^{\alpha}\psi\right]
 + m\bar\psi \psi\right\},\nonumber
\end{equation} 
%\end{footnotesize}
where we write the Lagrangian of matter as
\begin{equation}
\label{acao_espinorial_1}
 {\cal L}_M = \det(e) \left\{\frac{i}{2} \left[\bar\psi\gamma^{\alpha}\,\,{e_{\alpha}}^{\mu}
\nabla_{\mu}\psi -  {e_{\alpha}}^{\mu}
 (\nabla_{\mu}\bar{\psi})\gamma^{\alpha}\psi\right]
 + m\bar\psi \psi\right\},
\end{equation} 
being that $S_{M}=S_{M}(\psi,{e_{\alpha}}^{\mu},{\Gamma^{\gamma\delta}}_{\mu})$.
Let us then do an exercise to obtain the energy-momentum tensor for the spinor field, calculating the variation of the Lagrangian in relation to vierbein field,
\begin{equation}
\label{Tensor_E-P_2}
 \frac{\delta {\cal L}_{M}}{\delta {e_{\gamma}}^{\rho}}= \frac{\delta {\cal L}_{M}}{\delta g^{\mu\nu}}
 \,\, \frac{\delta g^{\mu\nu}}{\delta {e_{\gamma}}^{\rho}},
\end{equation}
where we can identify the term $\dfrac{\delta {\cal L}_{M}}{\delta g^{\mu\nu}}$ 
with the energy-momentum tensor,
$$\frac{\delta {\cal L}_{M}}{\delta g^{\mu\nu}} = \frac{\sqrt{-g}}{2}T_{\mu\nu}.$$
We need to calculate the term $\dfrac{\delta g^{\mu\nu}}{\delta {e_{\gamma}}^{\rho}} $. We can do it using the expression
$ g^{\mu\nu}={e_{\alpha}}^{\mu}{e_{\beta}}^{\nu}\eta^{\alpha\beta}$,
so that
\begin{equation}
 \delta g^{\mu\nu}=(\delta{e_{\alpha}}^{\mu})\,{e_{\beta}}^{\nu}\eta^{\alpha\beta}
 + {e_{\alpha}}^{\mu}(\delta{e_{\beta}}^{\nu})\eta^{\alpha\beta},
\end{equation}
so we can get
\begin{equation}
\frac{\delta g^{\mu\nu}}{\delta {e_{\gamma}}^{\rho}} =
\left(\frac{\delta{e_{\alpha}}^{\mu}}{\delta{e_{\gamma}}^{\rho}} \right){e_{\beta}}^{\nu}\eta^{\alpha\beta} +
\left(\frac{\delta{e_{\beta}}^{\nu}}{\delta{e_{\gamma}}^{\rho}} \right){e_{\alpha}}^{\mu}\eta^{\alpha\beta}
={\delta_{\alpha}}^{\gamma}{\delta_{\rho}}^{\mu}{e_{\beta}}^{\nu}\eta^{\alpha\beta} + 
{\delta_{\beta}}^{\gamma}{\delta_{\rho}}^{\nu}{e_{\alpha}}^{\mu}  \eta^{\alpha\beta}
={\delta_{\rho}}^{\mu}{e_{\beta}}^{\nu}\eta^{\beta\gamma} + 
{\delta_{\rho}}^{\nu}{e_{\alpha}}^{\mu}  \eta^{\alpha\gamma}.\nonumber
\end{equation}
Then, replacing these results in the expression (\ref{Tensor_E-P_2}), we have
\begin{equation}
  \frac{\delta {\cal L}_{M}}{\delta {e_{\gamma}}^{\rho}} =  \frac{\sqrt{-g}}{2}T_{\mu\nu}
  ({\delta_{\rho}}^{\mu}{e_{\beta}}^{\nu}\eta^{\beta\gamma} + 
{\delta_{\rho}}^{\nu}{e_{\alpha}}^{\mu}  \eta^{\alpha\gamma})
= \det(e)\,\,T_{\mu\nu}{\delta_{\rho}}^{\mu}{e_{\beta}}^{\nu}\eta^{\beta\gamma} 
= \det(e)\,\,T_{\rho\nu}{e_{\beta}}^{\nu}\eta^{\beta\gamma},\nonumber
\end{equation}
and multiplying the above equation by ${e_{\gamma}}^{\sigma}$ we obtain
\begin{equation}
{e_{\gamma}}^{\sigma} \frac{\delta {\cal L}_{M}}{\delta {e_{\gamma}}^{\rho}}= 
 \det(e)\,\,T_{\rho\nu}{e_{\beta}}^{\nu}{e_{\gamma}}^{\sigma}\eta^{\beta\gamma},\nonumber
\end{equation}
and with aid
$$ {e_{\beta}}^{\nu}{e_{\gamma}}^{\sigma}\eta^{\beta\gamma}= g^{\nu\sigma},$$
then we have
\begin{equation}
{e_{\gamma}}^{\sigma} \frac{\delta {\cal L}_{M}}{\delta {e_{\gamma}}^{\rho}}= 
 \det(e)\,\,T_{\rho\nu}g^{\nu\sigma},\nonumber
\end{equation}
which results in
\begin{equation}
\label{Tensor_E-P_espinor_1}
\frac{{e_{\gamma}}^{\sigma}}{\det(e)} \,\, \frac{\delta {\cal L}_{M}}{\delta {e_{\gamma}}^{\rho}}= 
 {T_{\rho}}^{\sigma},
\end{equation}
or 
\begin{equation}
\label{Tensor_E-P_espinor_2}
 T_{\mu\nu} = 
\frac{e_{\alpha\mu}}{\det(e)} \,\, \frac{\delta {\cal L}_{M}}{\delta {e_{\alpha}}^{\nu}} . 
\end{equation}

Let us then compute the energy-momentum tensor for the spinorial or Dirac field  given by the Lagrangian of the action (\ref{acao_espinorial_1}).
Using the fact that
\begin{equation}
 \delta[\det(e)]= -\det(e){\omega^{\alpha}}_{\mu}\delta{e_{\alpha}}^{\mu},
\end{equation}
so that $\dfrac{\delta {\cal L}_{M}}{\delta {e_{\alpha}}^{\nu}}$, follows that
\begin{equation}
 \delta {\cal L}_{M} =  -\det(e){\omega^{\alpha}}_{\sigma}\delta{e_{\alpha}}^{\sigma}
\left\{ \frac{i}{2} \left[\bar\psi\gamma^{\alpha}\,\,{e_{\alpha}}^{\rho}
\nabla_{\rho}\psi -  {e_{\alpha}}^{\rho}
 (\nabla_{\rho}\bar{\psi})\gamma^{\alpha}\psi\right] + m\bar\psi\psi \right\}
+ \det(e) \left\{ \frac{i}{2} \left[\bar\psi\gamma^{\alpha}\,\,(\delta{e_{\alpha}}^{\rho})
\nabla_{\rho}\psi -  (\delta{e_{\alpha}}^{\rho})
 (\nabla_{\rho}\bar{\psi})\gamma^{\alpha}\psi\right] \right\}.
 \end{equation}
The first term of the expression is a constraint term that can be written as follows,
\begin{equation}
 \left\{ \frac{i}{2} \left[\bar\psi\gamma^{\alpha}\,\,{e_{\alpha}}^{\rho}
\nabla_{\rho}\psi -  {e_{\alpha}}^{\rho}
 (\nabla_{\rho}\bar{\psi})\gamma^{\alpha}\psi\right]+ m\bar\psi\psi \right\} = 
 \frac{1}{2}\bar\psi\left[i \gamma^{\rho}\nabla_{\rho}\psi +m\psi\right] 
 - \frac{1}{2} [i \nabla_{\rho}\bar{\psi}\gamma^{\rho} - m\bar\psi]\psi. \nonumber
\end{equation}
We then see that the terms in square brackets satisfy the Dirac equations,
\begin{equation}
 i \gamma^{\rho}\nabla_{\rho}\psi + m\psi = 0
\hspace*{1cm} \mbox{and} \hspace*{1cm}
 i \nabla_{\rho}\bar{\psi}\gamma^{\rho} - m\bar\psi=0,
\end{equation}
so that we have
\begin{eqnarray}
 \frac{\delta {\cal L}_{M}}{\delta{e_{\alpha}}^{\nu}} &=&  
 \det(e) \left\{ \frac{i}{2} \left[\bar\psi\gamma^{\beta}\,\,
 \left(\frac{\delta{e_{\beta}}^{\rho}}{\delta{e_{\alpha}}^{\nu}}\right)
\nabla_{\rho}\psi -  \left(\frac{\delta{e_{\beta}}^{\rho}}{\delta{e_{\alpha}}^{\nu}}\right)
 (\nabla_{\rho}\bar{\psi})\gamma^{\beta}\psi\right] \right\}\cr
 &=& \det(e) \left\{ \frac{i}{2} \left[\bar\psi\gamma^{\beta}\,\,
 ({\delta_{\beta}}^{\alpha}{\delta_{\nu}}^{\rho})
\nabla_{\rho}\psi - ({\delta_{\beta}}^{\alpha}{\delta_{\nu}}^{\rho})
 (\nabla_{\rho}\bar{\psi})\gamma^{\beta}\psi\right] \right\}\cr
 &=&  \det(e) \left\{ \frac{i}{2} \left[\bar\psi\gamma^{\alpha}
\nabla_{\nu}\psi -  (\nabla_{\nu}\bar{\psi})\gamma^{\alpha}\psi\right] \right\},\nonumber
 \end{eqnarray}
then the energy-momentum tensor (\ref{Tensor_E-P_espinor_2}) becomes
\begin{equation}
 T_{\mu\nu} = 
\frac{e_{\alpha\mu}}{\det(e)} \,\, \frac{\delta {\cal L}_{M}}{\delta {e_{\alpha}}^{\nu}}=
\frac{e_{\alpha\mu}}{\det(e)}\det(e) \left\{ \frac{i}{2} \left[\bar\psi\gamma^{\alpha}
\nabla_{\nu}\psi -  (\nabla_{\nu}\bar{\psi})\gamma^{\alpha}\psi\right] \right\}\nonumber 
\end{equation}
or
\begin{equation}
\label{Tensor_E-P_espinor_3}
 T_{\mu\nu} =  \frac{i}{2} \left[\bar\psi\gamma_{\mu}
\nabla_{\nu}\psi -  (\nabla_{\nu}\bar{\psi})\gamma_{\mu}\psi\right]. 
\end{equation}

The energy-momentum tensor must be symmetrical. In the section \ref{campo_espinorial}, we have seen that in the flat spacetime, in the Classical Theory of Fields, the calculation of the energy-momentum tensor obtained by the Noether theorem results in an energy-momentum tensor similar to that obtained in the above equation (\ref{Tensor_E-P_espinor_3}) 
and we have seen how we can use the mechanism of Belifante-Rosenfeld to symmetry the  energy-momentum tensor. We have seen that the contribution of spin current density contributes to the energy-momentum tensor. Here the same problem occurs, the spin current density contributes to the mathematical expression of the Hilbert energy-momentum tensor. Let us see how to perform these calculations, first note that the variation of the action of matter can be rewritten as,
\begin{eqnarray}
\label{acao_EC_5}
\delta S_{M} = \frac{1}{2} \int_{\Omega} d^4x  \sqrt{-g}\,\, T^{\mu\nu} \delta g_{\mu\nu}  -   \frac{1}{2} \int_{\Omega} d^4x \sqrt{-g}\,\,  {{\mathfrak{S}^{\nu}}_{\rho}}^{\mu}\delta{\Gamma^{\rho}}_{\mu\nu}. 
\end{eqnarray}
where we use equation (\ref{equacao_de_campo_Einstein_1}),
$G_{\mu\nu} = 8\pi G\, T_{\mu\nu}$ and also the equation (\ref{equacao_de_campo_Cartan_2})
$ {\mathsf T}_{\mu\lambda\nu}  +  {\mathsf T}_{\nu}g_{\lambda\mu}  - {\mathsf T}_{\lambda}g_{\mu\nu} =  8\pi G \,\mathfrak{S}_{\nu\lambda\mu}$.
The term $ \delta{\Gamma^{\rho}}_{\mu\nu}$ in the second integral of the above equation  (\ref{acao_EC_5}) can be rewritten in terms of $\delta g_{\mu\nu}$. 
Let us perform this calculation from the identity,
\begin{equation}
 \delta(\nabla_{\nu} g_{\mu\lambda}) = 0,
\end{equation}
where we can obtain
\begin{equation}
 \partial_{\nu}\delta g_{\mu\lambda} - (\delta g_{\mu\rho}){\Gamma^{\rho}}_{\nu\lambda} - (\delta g_{\lambda\rho}){\Gamma^{\rho}}_{\nu\mu} - g_{\mu\rho} \delta {\Gamma^{\rho}}_{\nu\lambda} -  g_{\lambda\rho}\delta {\Gamma^{\rho}}_{\nu\mu} = 0.\nonumber
\end{equation}
Just as we assemble a system of three equations seen in the equation (\ref{sistema_1}), let us put together a system with the permutations of the indices in the above equation,
\begin{equation}
 \begin{cases}
  \nabla_{\nu}(\delta g_{\mu\lambda}) =  g_{\mu\rho} \delta {\Gamma^{\rho}}_{\nu\lambda} +  g_{\lambda\rho}\delta {\Gamma^{\rho}}_{\nu\mu}\cr
  \nabla_{\mu}(\delta g_{\nu\lambda}) =  g_{\nu\rho} \delta {\Gamma^{\rho}}_{\mu\lambda} +  g_{\lambda\rho}\delta {\Gamma^{\rho}}_{\mu\nu}\cr
  \nabla_{\lambda}(\delta g_{\mu\nu}) =  g_{\mu\rho} \delta {\Gamma^{\rho}}_{\lambda\nu} +  g_{\nu\rho}\delta {\Gamma^{\rho}}_{\lambda\mu}
 \end{cases},
\end{equation}
and then add the first two equations and subtract the last one so that we will have,
\begin{equation}
 \nabla_{\nu}(\delta g_{\mu\lambda})+ \nabla_{\mu}(\delta g_{\nu\lambda}) - \nabla_{\lambda}(\delta g_{\mu\nu}) =  g_{\lambda\rho}(\delta {\Gamma^{\rho}}_{\mu\nu} + \delta {\Gamma^{\rho}}_{\nu\mu}) +  g_{\mu\rho}( \delta {\Gamma^{\rho}}_{\nu\lambda} - \delta {\Gamma^{\rho}}_{\lambda\nu}) + g_{\nu\rho} (\delta {\Gamma^{\rho}}_{\mu\lambda} -  \delta {\Gamma^{\rho}}_{\lambda\mu}), \nonumber
\end{equation}
where we should use the definition of the torsion tensor (\ref{tensor_torcao_1}), 
$ {{\mathsf T}^{\kappa}}_{\lambda\mu}  = {\Gamma^{\kappa}}_{\lambda\mu}  - {\Gamma^{\kappa}}_{\mu\lambda}$, so we get
\begin{equation}
   g_{\lambda\rho}[\delta {\Gamma^{\rho}}_{\mu\nu} + (\delta {\Gamma^{\rho}}_{\mu\nu}- \delta {\mathsf T ^{\rho}}_{\mu\nu})] =\nabla_{\nu}(\delta g_{\mu\lambda})+ \nabla_{\mu}(\delta g_{\nu\lambda}) - \nabla_{\lambda}(\delta g_{\mu\nu})  +  g_{\mu\rho} \delta {\mathsf T ^{\rho}}_{\lambda\nu}  + g_{\nu\rho} \delta {\mathsf T^{\rho}}_{\lambda\mu} ,\nonumber
\end{equation}
which then results in the equation
\begin{equation}
  \delta {\Gamma^{\rho}}_{\mu\nu}   = \frac{1}{2} g^{\lambda\rho} [\nabla_{\nu}(\delta g_{\mu\lambda})+ \nabla_{\mu}(\delta g_{\nu\lambda}) - \nabla_{\lambda}(\delta g_{\mu\nu}]  +\frac{1}{2}g^{\lambda\rho}[  g_{\mu\kappa} \delta {\mathsf T ^{\kappa}}_{\lambda\nu}  + g_{\nu\kappa} \delta {\mathsf T^{\kappa}}_{\lambda\mu} + g_{\lambda\kappa} \delta {\mathsf T ^{\kappa}}_{\mu\nu}].\nonumber
\end{equation}
Now let us replace this expression above in the second term of the action variation in equation (\ref{acao_EC_5}),
\begin{eqnarray}
\label{acao_EC_6}
\frac{1}{2} \int_{\Omega} d^4x \sqrt{-g}\,\,  {{\mathfrak{S}^{\nu}}_{\rho}}^{\mu}\delta{\Gamma^{\rho}}_{\mu\nu} &=& \frac{1}{4} \int_{\Omega} d^4x \sqrt{-g}\,\,  {{\mathfrak{S}^{\nu}}_{\rho}}^{\mu}  g^{\lambda\rho} [\nabla_{\nu}(\delta g_{\mu\lambda})+ \nabla_{\mu}(\delta g_{\nu\lambda}) - \nabla_{\lambda}(\delta g_{\mu\nu}]\cr
& &+ \frac{1}{4} \int_{\Omega} d^4x \sqrt{-g}\,\,  {{\mathfrak{S}^{\nu}}_{\rho}}^{\mu} g^{\lambda\rho}[  g_{\mu\kappa} \delta {\mathsf T ^{\kappa}}_{\lambda\nu}  + g_{\nu\kappa} \delta {\mathsf T^{\kappa}}_{\lambda\mu} + g_{\lambda\kappa} \delta {\mathsf T ^{\kappa}}_{\mu\nu}]\cr
&=& \frac{1}{4} \int_{\Omega} d^4x \sqrt{-g}\,\,  {\mathfrak{S}^{\nu\lambda\mu}} [\nabla_{\nu}(\delta g_{\mu\lambda})+ \nabla_{\mu}(\delta g_{\nu\lambda}) - \nabla_{\lambda}(\delta g_{\mu\nu}]\cr
& & + \frac{1}{4} \int_{\Omega} d^4x \sqrt{-g}\,[ {\mathfrak{S}^{\nu\lambda}}_{\kappa}\delta {\mathsf T ^{\kappa}}_{\lambda\nu} + {{\mathfrak{S}_{\kappa}}}^{\lambda\mu} \delta {\mathsf T^{\kappa}}_{\lambda\mu} + {{\mathfrak{S}^{\nu}}_{\kappa}}^{\mu}\delta {\mathsf T ^{\kappa}}_{\mu\nu}]. 
\end{eqnarray}
The second integral above can be rewritten as follows below,
\begin{equation}
 \frac{1}{4} \int_{\Omega} d^4x \sqrt{-g}\,[ {\mathfrak{S}^{\nu\lambda}}_{\kappa}\delta {\mathsf T ^{\kappa}}_{\lambda\nu} + {{\mathfrak{S}_{\kappa}}}^{\lambda\mu} \delta {\mathsf T^{\kappa}}_{\lambda\mu} + {{\mathfrak{S}^{\nu}}_{\kappa}}^{\mu}\delta {\mathsf T ^{\kappa}}_{\mu\nu}] = 
 \frac{1}{4} \int_{\Omega} d^4x \sqrt{-g}\, {\mathfrak{S}}^{\nu\rho\mu}( \delta {\mathsf T}_{\mu\rho\nu} + \delta {\mathsf T}_{\nu\rho\mu} + \delta {\mathsf T}_{\rho\mu\nu}), \nonumber
\end{equation}
where the term enclosed in parenthesis on the right hand side of above equation is the variation of the contorsion tensor (\ref{contorcao}), 
and which can be terminated in the below equation,
\begin{equation}
 \frac{1}{4} \int_{\Omega} d^4x \sqrt{-g}\,[ {\mathfrak{S}^{\nu\lambda}}_{\kappa}\delta {\mathsf T ^{\kappa}}_{\lambda\nu} + {{\mathfrak{S}_{\kappa}}}^{\lambda\mu} \delta {\mathsf T^{\kappa}}_{\lambda\mu} + {{\mathfrak{S}^{\nu}}_{\kappa}}^{\mu}\delta {\mathsf T ^{\kappa}}_{\mu\nu}] = 
 \frac{1}{4} \int_{\Omega} d^4x \sqrt{-g}\, {\mathfrak{S}}^{\nu\rho\mu} \delta K_{\rho\mu\nu}. \nonumber
\end{equation}

Now let us take the first term from the first part of the integral (\ref{acao_EC_6}), ${\mathfrak{S}^{\nu\lambda\mu}} \nabla_{\nu}(\delta g_{\mu\lambda}) $, 
to calculate a integral by parts. Let us see that,
\begin{equation}
\label{acao_EC_7}
  \frac{1}{4} \int_{\Omega} d^4x \sqrt{-g}\,{\mathfrak{S}^{\nu\lambda\mu}} \nabla_{\nu}(\delta g_{\mu\lambda}) =  \frac{1}{4} \int_{\Omega} d^4x \sqrt{-g}\,[\nabla_{\nu} ({\mathfrak{S}^{\nu\lambda\mu}} \delta g_{\mu\lambda}) - (\nabla_{\nu}{\mathfrak{S}^{\nu\lambda\mu}}) \delta g_{\mu\lambda}].
\end{equation}
Let us use the fact that a total divergence integral in a Classical Fields Theory in a spacetime with torsion results in the expression (\ref{divergencia_2}) where,
\begin{equation}
\int_{\Omega} d^4x \sqrt{-g}\, \nabla_{\mu} V^{\mu} = - \int_{\Omega} d^4x \sqrt{-g}\, {\mathsf T}_{\mu} V^{\mu} . \nonumber
\end{equation}
Using this result in the integral (\ref{acao_EC_7}) we have
\begin{equation}
\label{acao_EC_8}
  \frac{1}{4} \int_{\Omega} d^4x \sqrt{-g}\,{\mathfrak{S}^{\nu\lambda\mu}} \nabla_{\nu}(\delta g_{\mu\lambda}) =  \frac{1}{4} \int_{\Omega} d^4x \sqrt{-g}\,[-{\mathsf T}_{\nu} ({\mathfrak{S}^{\nu\lambda\mu}} \delta g_{\mu\lambda}) - (\nabla_{\nu}{\mathfrak{S}^{\nu\lambda\mu}}) \delta g_{\mu\lambda}].
\end{equation}
Using this result above and substituting in equation (\ref{acao_EC_6}) we obtain
\begin{eqnarray}
\label{acao_EC_9}
\frac{1}{2} \int_{\Omega} d^4x \sqrt{-g}\,\,  {{\mathfrak{S}^{\nu}}_{\rho}}^{\mu}\delta{\Gamma^{\rho}}_{\mu\nu}
&=&  - \frac{1}{4} \int_{\Omega} d^4x \sqrt{-g}\,\nabla_{\lambda}( {\mathfrak{S}^{\mu\nu\lambda}} + {\mathfrak{S}^{\lambda\mu\nu}}- {\mathfrak{S}^{\nu\lambda\mu}})\delta g_{\mu\nu}\cr
& &- \frac{1}{4} \int_{\Omega} d^4x \sqrt{-g}\,{\mathsf T}_{\lambda}( {\mathfrak{S}^{\mu\nu\lambda}} + {\mathfrak{S}^{\lambda\mu\nu}}- {\mathfrak{S}^{\nu\lambda\mu}})\delta g_{\mu\nu}
+ \frac{1}{4} \int_{\Omega} d^4x \sqrt{-g}\, {\mathfrak{S}}^{\nu\rho\mu} \delta K_{\rho\mu\nu}. \nonumber
\end{eqnarray}
The second term of the integral above cancels due to antisymmetry in the indices $\mu$ and $\nu$ 
in the spin current tensor density contracted with the variation of the symmetrical metric tensor. Turning these values into equation (\ref{acao_EC_5}) it follows that,
\begin{eqnarray}
\label{acao_EC_10}
\delta S_{M} = \frac{1}{2} \int_{\Omega} d^4x  \sqrt{-g}\,\, \left [T^{\mu\nu}   +   \frac{1}{2} \nabla_{\lambda}( {\mathfrak{S}^{\mu\nu\lambda}} + {\mathfrak{S}^{\lambda\mu\nu}}- {\mathfrak{S}^{\nu\lambda\mu}})\right]\delta g_{\mu\nu} - \frac{1}{4} \int_{\Omega} d^4x \sqrt{-g}\, {\mathfrak{S}}^{\nu\rho\mu} \delta K_{\rho\mu\nu}. 
\end{eqnarray}
Then one must observe closely the term between brackets as the symmetric energy-momentum tensor of Belinfante-Rosenfeld,
\begin{equation}
\label{tensor_Belinfante-Rosenfeld_2}
 {\mathfrak T }^{\mu\nu} = T^{\mu\nu}   +   \frac{1}{2} \nabla_{\lambda}( {\mathfrak{S}^{\mu\nu\lambda}} + {\mathfrak{S}^{\lambda\mu\nu}}- {\mathfrak{S}^{\nu\lambda\mu}})
\end{equation}
for a curved spacetime. Comparing the equation above with the Belifante-Rosenfeld tensor equation obtained in equation (\ref{tensor_Belinfante-Rosenfeld}) 
through the Noether theorem in Classical Fields Theory in a flat spacetime, we see that the density of spin current tensor is given by,
\begin{equation}
 {\mathfrak{S}^{\lambda\mu\nu}} = -i S^{\lambda\mu\nu}.
\end{equation}
With aid of equation (\ref{densidade_spin_3}) we can identify the spin current density tensor as
\begin{equation}
\label{densidade_spin_4}
{\mathfrak{S}_{\lambda\mu\nu}} = \frac{1}{2} \,\bar\psi \left\{\gamma_{\lambda},\Sigma_{\mu\nu}\right\}\psi,
\hspace*{1cm}\mbox{or else} \hspace*{1cm} {\mathfrak{S}_{\lambda\mu\nu}} = \frac{i}{8} \,\bar\psi \left\{\gamma_{\lambda},[\gamma_{\mu},\gamma_{\nu}] \right\}\psi.
\end{equation}
Performing the same calculation procedures as in section \ref{campo_espinorial} and replacing equation (\ref{Tensor_E-P_espinor_3}) into (\ref{tensor_Belinfante-Rosenfeld_2}) we obtain that the energy-momentum tensor symmetrical for the spinorial field is given by  
\begin{equation}
\label{Tensor_E-P_espinor_4}
 {\mathfrak T}_{\mu\nu} =\frac{i}{4} \left[\bar\psi\gamma_{\mu} \nabla_{\nu}\psi + \bar\psi\gamma_{\nu} \nabla_{\mu}\psi -  ({\nabla_{\mu}\bar\psi})\gamma_{\nu}\psi - ({\nabla_{\nu}\bar\psi})\gamma_{\mu}\psi \right].
\end{equation}
It is the energy-momentum tensor that generates the curvature of spacetime, given by the field equation (\ref{equacao_de_campo_Einstein_1}), 
emphasizing that the energy-momentum tensor above contains the density of spin current, as we have seen in the previous details.

Now let us look at the second equation of motion that relates the  density of spin current tensor to the torsion in spacetime. A variation in spinorial action (\ref{acao_Dirac_gravitacao_2}) 
in relation to connection ${\Gamma^{\alpha}}_{\beta\gamma}$ it results in
\begin{eqnarray}
\label{acao_Dirac_gravitacao_3}
 \frac{\delta S_{M}}{\delta {\Gamma^{\alpha}}_{\beta\gamma}} &=& \frac{\delta}{\delta {\Gamma^{\alpha}}_{\beta\gamma}}\int_{\Omega}d^4 x\,\,\det(e) \left\{\frac{i}{2} \left[\bar\psi\gamma^{\zeta}\,\,{e_{\zeta}}^{\mu}
\left(\partial_{\mu}+ 
 \frac{i}{2}\Gamma_{\delta\mu\epsilon}\Sigma^{\delta\epsilon}\right)\psi -  {e_{\zeta}}^{\mu}
 \left(\partial_{\mu}\bar\psi- 
 \frac{i}{2} \Gamma_{\delta\mu\epsilon} \bar\psi\,\Sigma^{\delta\epsilon}\right)\gamma^{\zeta}\psi\right]
 + m\bar\psi \psi\right\}\cr\cr
&=& \left(\frac{i}{2}\right)^2\,\det(e)\, \left(\bar\psi\gamma^{\zeta} \frac{\delta \Gamma_{\delta\zeta\epsilon}}{\delta {\Gamma^{\alpha}}_{\beta\gamma}}\Sigma^{\delta\epsilon}\psi + \bar\psi \frac{\delta \Gamma_{\delta\zeta\epsilon}}{\delta {\Gamma^{\alpha}}_{\beta\gamma}}\Sigma^{\delta\epsilon}\gamma^{\zeta}\psi\right)
\hspace*{6pt} = \hspace*{6pt} -\frac{\det(e)}{4}\,\,\bar\psi \left\{\gamma^{\beta},{\Sigma_{\alpha}}^{\gamma} \right\}\psi, \nonumber
\end{eqnarray} 
now comparing the above result with equation (\ref{densidade_spin_4}) we arrive at,
\begin{equation}
 \frac{\delta S_{M}}{\delta {\Gamma^{\alpha}}_{\beta\gamma}} = -\frac{\det(e)}{2} {{{\mathfrak{S}^{\beta}}_{\alpha}}}^{\gamma},
\end{equation}
or expressing the spin current density tensor as
\begin{equation}
 \mathfrak{S}_{\beta\alpha\gamma} = -\frac{2}{\det(e)} \frac{\delta S_{M}}{\delta \Gamma^{\alpha\beta\gamma}}.
\end{equation}
We must compare the above expression with the definition in equation (\ref{equacao_de_campo_Cartan_3}), where we have the second Einstein-Cartan equation, where the $U_4$ spacetime torsion originates from the spin current density tensor, given by the equation below in Minkowski orthornormal coordinates,
\begin{equation}
 \label{equacao_de_campo_Cartan_4}
  {\mathsf T}_{\alpha\gamma\beta}  +  {\mathsf T}_{\gamma}\eta_{\alpha\beta}  - {\mathsf T}_{\beta}\eta_{\alpha\gamma} =  8\pi G \,\mathfrak{S}_{\alpha\beta\gamma}.
\end{equation}

%%%%%%%%%%%%%%%%%%%%%%%%%%%%%%%%%%%%%%%%%%%%%%%%%%%%%%%%%%%%%%%%%%%%%%%%%%%%%%%%%%%%%%%%%%%%%%%%%%

\section{Conclusion}

The Einstein-Cartan-Sciama-Kibble theory described in the Riemann-Cartan spacetime, $ U_4 $, is obtained by the Gauge Theory under local transformations of Poincar\'e in the field of spinorial matter. The fundamental constituents of matter are fermions (spin $\frac{1}{2}$), so that the Einstein-Cartan-Sciama-Kibble theory is obtained by gauge of the action of the spinorial field.

Under the global invariance of Poincar\'e transformations the energy-momentum tensor is conserved and the angular momentum current is conserved. We have seen in the section \ref{campo_espinorial} that in the absence of gravity, in Minkowski spacetime, through the Noether theorem the energy-momentum tensor is not symmetric, and it is necessary the Belinfante-Rosenfeld procedures that add to the energy-momentum tensor the contributions of the spin current density of the fermionic field, which makes the energy-momentum tensor symmetrical. The requirement that the energy-momentum tensor be symmetric comes from the requirement in which the metric energy-momentum tensor  or Hilbert energy-momentum tensor is symmetric by definition according to equation (\ref{Tensor_E-P_1}).

The spinor action under local transformations of Poincar\'e brings the interaction with gravitational fields. The starting point must be the action (\ref{acao_Dirac_gravitacao_2}), valid in Minkowski  flat spacetime when the tetrad field reduces in ${e_{\alpha}}^{\mu} ={\delta_{\alpha}}^{\mu}$, ${\omega^{\beta}}_{\nu} = {\delta^{\beta}}_{\nu}  $, $\det(e)=1$ and the connection $\Gamma_{\alpha\mu\beta}=0$. The tetrad field ${e_{\alpha}}^{\mu}$ maps the curved spacetime, non-inertial system, in a locally flat spacetime, inertial system, by equation (\ref{vierbeins}). Then an observer in another system of references observing the spinorial field, will make the observation under local transformations of Poincar\'e getting the emergence of two gauge fields: ${e_{\alpha}}^{\mu}$ and $\Gamma_{\alpha\mu\beta}$. The covariant derivative (\ref{del_covariante_espinor_5}) 
is the translation generator of the Poincar\'e group $P(1,3)$ and $\Sigma_{\alpha\beta}$ are the Lorentz rotation and boost generators \cite{Hehl_1, Hehl_3}. The local invariance of Poincar\'e leads to the field equation (\ref{equacao_de_campo_Einstein_1}) 
where in the same way as in General Relavity theory, the energy-momentum tensor is the source of the curvature. The other field equation obtained (\ref{equacao_de_campo_Cartan_4}) shows that the density of spin current tensor becomes the source of torsion in spacetime $U_4$.

It should be noted that in Einstein General Relativity the tensor curvature is calculated through derivatives in the connections which are only the Christoffel symbols, which results in second-order differential equations in the metric tensor $g_{\mu\nu}$ implying that gravitational interaction propagates in Riemannian spacetime. The same happens in the spacetime of Riemann-Cartan that can have non-zero torsion.
With this in mind we should note that equation (\ref{equacao_de_campo_Cartan_4}) 
which relates torsion to spin current density tensor is not a differential equation such as the field equation (\ref{equacao_de_campo_Einstein_1}) but instead is an algebraic relationship between torsion and spin current density tensor. This implies that the Riemann-Cartan spacetime gravitation will have nonzero torsion only in regions where fermionic matter exists. The torsion in the Riemann-Cartan spacetime can not be dissociated from the fermionic matter and consequently can not propagate in the vacuum as a torsion wave or through another interaction \cite{Hehl_4}.

The Einstein-Cartan theory of gravitation in spacetime with torsion adds to gravitation the existence of a weak interaction between the gravitational field and the fermionic matter. Calculations and discussions have shown that the density of matter containing fermions with intrinsic angular momentum in units of $\frac{\hbar}{2}$, must be of the order of $10^{47}$ g/cm$^3$ 
for electrons and $10^{54}$ g/cm$^3$ 
for neutrons, so that there was the possibility of estimating significant deviations from the predictions of General Relativity. To have an idea of these dimensions, compare the matter density of a neutron star that is of the order of $10^{16}$ g/cm$^3$  \cite{Hehl_1}.Certainly for such high densities, the expected effects that fermionic matter can twist the spacetime must be for extreme conditions in the gravitational collapses approached in cosmology and in the big bang. There are also expectations of such spin-torsion effects occurring on the Planck scale where quantum gravity processes begin to be relevant \cite{Hehl_1,Hehl_4}.

%%%%%%%%%%%%%%%%%%%%%%%%%%%%%%%%%%%%%%%%%%%%%%%%%%%%%%%%%%%%%%%%%%%%%%%%%%%%%%%%%%%%%

%%%%%%%%%%%%%%%%%%%%%%%%%%%%%%%%%%%%%%%%%%%%%%%%%%%%%%%%%%%%%%%%%%%%%%%%%%%%%%%%%%%%%

%%%%%%%%%%%%%%%%%%%%%%%%%%%%%%%%%%%%%%%%%%%%%%%%%%%%%%%%%%%%%%%%%%%%%%%%%%%%%%%%%%%%%

%%%%%%%%%%%%%%%%%%%%%%%%%%%%%%%%%%%%%%%%%%%%%%%%%%%%%%%%%%%%%%%%%%%%%%%%%%%%%%%%%%%%%

%%%%%%%%%%%%%%%%%%%%%%%%%%%%%%%%%%%%%%%%%%%%%%%%%%%%%%%%%%%%%%%%%%%%%%%%%%%%%%%%%%%%%

%%%%%%%%%%%%%%%%%%%%%%%%%%%%%%%%%%%%%%%%%%%%%%%%%%%%%%%%%%%%%%%%%%%%%%%%%%%%%%%%%%%%%

%%%%%%%%%%%%%%%%%%%%%%%%%%%%%%%%%%%%%%%%%%%%%%%%%%%%%%%%%%%%%%%%%%%%%%%%%%%%%%%%%%%%%

%%%%%%%%%%%%%%%%%%%%%%%%%%%%%%%%%%%%%%%%%%%%%%%%%%%%%%%%%%%%%%%%%%%%%%%%%%%%%%%%%%%%%

%%%%%%%%%%%%%%%%%%%%%%%%%%%%%%%%%%%%%%%%%%%%%%%%%%%%%%%%%%%%%%%%%%%%%%%%%%%%%%%%%%%%%

%%%%%%%%%%%%%%%%%%%%%%%%%%%%%%%%%%%%%%%%%%%%%%%%%%%%%%%%%%%%%%%%%%%%%%%%%%%%%%%%%%%%%

\appendix
\section{Commutator of covariant derivatives}

Let be the covariant derivative (\ref{del_covariante_espinor_5})
that leaves the spinorial action  (\ref{acao_Dirac_gravitacao_2}) 
invariant by local coordinate changes and Lorentz transformations, we can calculate the gauge curvature given by
$ [\nabla_{\alpha},\nabla_{\beta}] $. Therefore,
\begin{equation}
\label{Operador_curvatura_1A}
 [\nabla_{\alpha},\nabla_{\beta}] = \nabla_{\alpha}\nabla_{\beta} - \nabla_{\beta}\nabla_{\alpha}
  = \nabla_{\alpha} \left[{e_{\beta}}^{\nu}\left(\partial_{\nu}+ 
 \frac{i}{2}{{\Gamma^{\gamma}}_{\mu}}^{\delta}\Sigma_{\gamma\delta}\right)\right] - 
 \nabla_{\beta} \left[{e_{\alpha}}^{\mu}\left(\partial_{\mu}+ 
 \frac{i}{2}{{\Gamma^{\epsilon}}_{\mu}}^{\zeta}\Sigma_{\epsilon\zeta}\right)\right],
\end{equation}
with $\nabla_{\alpha}={e_{\alpha}}^{\mu}\nabla_{\mu}$ we have,
\begin{equation}
\label{Operador_curvatura_1}
 \nabla_{\alpha}\nabla_{\beta} = {e_{\alpha}}^{\mu}\nabla_{\mu}({e_{\beta}}^{\nu}\nabla_{\nu})
 = {e_{\alpha}}^{\mu}(\nabla_{\mu}{e_{\beta}}^{\nu} )\nabla_{\nu} + 
 {e_{\alpha}}^{\mu}{e_{\beta}}^{\nu}\nabla_{\mu}\nabla_{\nu} .
\end{equation}
Let us first look at the term ${e_{\alpha}}^{\mu}(\nabla_{\mu}{e_{\beta}}^{\nu} )\nabla_{\nu} $ 
of the above expression
\begin{equation}
\label{appendix_3}
{e_{\alpha}}^{\mu}\nabla_{\mu}{e_{\beta}}^{\nu} = {e_{\alpha}}^{\mu}\left(\partial_{\mu}+ 
 \frac{i}{2}{{\Gamma^{\epsilon}}_{\mu}}^{\zeta}\Sigma_{\epsilon\zeta}\right){e_{\beta}}^{\nu}
 = \partial_{\alpha} {e_{\beta}}^{\nu} +  
 \frac{i}{2}{{\Gamma^{\epsilon}}_{\alpha}}^{\zeta}\Sigma_{\epsilon\zeta}\,\,
 {e_{\beta}}^{\nu}.
\end{equation}

Then the Lorentz transformation generator of the group $SO(3,1)$ 
is acting on the vector index covariant $\beta$ of vierbein ${e_{\beta}}^{\nu}$, it follows that
\begin{equation}
 \Sigma_{\gamma\delta}\,\,{e_{\beta}}^{\nu} \rightarrow [\Sigma_{\gamma\delta}]^{\epsilon}_{\beta}
 \,\,{e_{\epsilon}}^{\nu}  
 = i(\eta_{\delta\beta}{\delta_{\gamma}}^{\epsilon} - \eta_{\gamma\beta}{\delta_{\delta}}^{\epsilon})
 {e_{\epsilon}}^{\nu} 
 =   i\eta_{\delta\beta} {e_{\gamma}}^{\nu}-i\eta_{\gamma\beta} {e_{\delta}}^{\nu} ,\nonumber
\end{equation}
which results for the equation (\ref{appendix_3}) the following expression
\begin{equation}
 \nabla_{\alpha}{e_{\beta}}^{\nu} =
\partial_{\alpha}{e_{\beta}}^{\nu} +  \frac{i}{2}{{\Gamma^{\gamma}}_{\alpha}}^{\delta} 
 ( i\eta_{\delta\beta} {e_{\gamma}}^{\nu}-i\eta_{\gamma\beta} {e_{\delta}}^{\nu})
= \partial_{\alpha} {e_{\beta}}^{\nu} -  \frac{1}{2}{\Gamma^{\gamma}}_{\alpha\beta}  {e_{\gamma}}^{\nu}+
 \frac{1}{2}{\Gamma_{\beta\alpha}}^{\delta}  {e_{\delta}}^{\nu} 
\nonumber
\end{equation} 
and with the algebraic property of connection non-coordinate basis $\Gamma_{\beta\alpha\gamma} = -\Gamma_{\alpha\beta\gamma}$, therefore the above equation results in 
\begin{equation}
\label{Operador_curvatura_3}
\nabla_{\alpha}{e_{\beta}}^{\nu} = \partial_{\alpha}{e_{\beta}}^{\nu} -
{e_{\gamma}}^{\nu}\,{\Gamma^{\gamma}}_{\alpha\beta} .
\end{equation}

Now we can return to the calculation of the equation (\ref{Operador_curvatura_1}) where it is
\begin{eqnarray}
%\label{Operador_curvatura_1}
 \nabla_{\alpha}\nabla_{\beta} = {e_{\alpha}}^{\mu}(\nabla_{\mu}{e_{\beta}}^{\nu})\nabla_{\nu} + 
 {e_{\alpha}}^{\mu}{e_{\beta}}^{\nu}\nabla_{\mu}\nabla_{\nu}, \nonumber
\end{eqnarray}
where we can substitute the result (\ref{Operador_curvatura_3}) in the first term and also replacing
$\nabla_{\nu}={\omega^{\gamma}}_{\nu}\nabla_{\gamma}$, we obtain
\begin{equation}
 \label{Operador_curvatura_6}
 \nabla_{\alpha}\nabla_{\beta} = {e_{\alpha}}^{\mu}( \partial_{\mu}{e_{\beta}}^{\nu} -
{e_{\delta}}^{\nu}\,{\Gamma^{\delta}}_{\mu\beta}) {\omega^{\gamma}}_{\nu} \nabla_{\gamma}+ 
 {e_{\alpha}}^{\mu}{e_{\beta}}^{\nu}\nabla_{\mu}\nabla_{\nu},
\end{equation}
where we must explain the term $\nabla_{\mu}\nabla_{\nu}$, where
\begin{equation}
 \nabla_{\mu}\nabla_{\nu} = \left(\partial_{\mu}+\frac{i}{2}{\Gamma^{\gamma\delta}}_{\mu}\Sigma_{\gamma\delta}\right)
 \left(\partial_{\nu}+\frac{i}{2}{\Gamma^{\epsilon\zeta}}_{\nu}\Sigma_{\epsilon\zeta}\right)
 = \partial_{\mu}\partial_{\nu}+\frac{i}{2}\partial_{\mu}{\Gamma^{\epsilon\zeta}}_{\nu}\Sigma_{\epsilon\zeta}
+\frac{i}{2}{\Gamma^{\epsilon\zeta}}_{\nu}\Sigma_{\epsilon\zeta} \partial_{\mu}
 +\frac{i}{2}{\Gamma^{\gamma\delta}}_{\mu}\Sigma_{\gamma\delta}\partial_{\nu}
  -\frac{1}{4}{\Gamma^{\gamma\delta}}_{\mu}
 {\Gamma^{\epsilon\zeta}}_{\nu}\Sigma_{\gamma\delta}\Sigma_{\epsilon\zeta},\nonumber
\end{equation}
so that the expression (\ref{Operador_curvatura_6}) becomes
\begin{equation}
\label{Operador_curvatura_7}
 \nabla_{\alpha}\nabla_{\beta} = {e_{\alpha}}^{\mu} {\omega^{\gamma}}_{\nu}( \partial_{\mu}{e_{\beta}}^{\nu} -
{e_{\delta}}^{\nu}\,{\Gamma^{\delta}}_{\mu\beta}) \nabla_{\gamma}
+
 {e_{\alpha}}^{\mu}{e_{\beta}}^{\nu}\big(
  \partial_{\mu}\partial_{\nu}+\frac{i}{2}\partial_{\mu}{\Gamma^{\epsilon\zeta}}_{\nu}\Sigma_{\epsilon\zeta}
+ \frac{i}{2}{\Gamma^{\gamma\delta}}_{\nu}\Sigma_{\gamma\delta} \partial_{\mu}
 +\frac{i}{2}{\Gamma^{\gamma\delta}}_{\mu}\Sigma_{\gamma\delta}\partial_{\nu}
  -\frac{1}{4}{\Gamma^{\gamma\delta}}_{\mu}
 {\Gamma^{\epsilon\zeta}}_{\nu}\Sigma_{\gamma\delta}\Sigma_{\epsilon\zeta}\big).
\end{equation}
And now we write the term $\nabla_{\beta}\nabla_{\alpha}$ from equation (\ref{Operador_curvatura_1A}), where we have
\begin{equation}
\label{Operador_curvatura_8}
 \nabla_{\beta}\nabla_{\alpha} = {e_{\beta}}^{\mu} {\omega^{\gamma}}_{\nu}( \partial_{\mu}{e_{\alpha}}^{\nu} -
{e_{\delta}}^{\nu}\,{\Gamma^{\delta}}_{\mu\alpha}) \nabla_{\gamma}
+
 {e_{\beta}}^{\mu}{e_{\alpha}}^{\nu}\big(
  \partial_{\mu}\partial_{\nu}+\frac{i}{2}{\partial_{\mu}\Gamma^{\epsilon\zeta}}_{\nu}\Sigma_{\epsilon\zeta}
+\frac{i}{2}{\Gamma^{\gamma\delta}}_{\nu}\Sigma_{\gamma\delta} \partial_{\mu}
 +\frac{i}{2}{\Gamma^{\gamma\delta}}_{\mu}\Sigma_{\gamma\delta}\partial_{\nu}
  -\frac{1}{4}{\Gamma^{\gamma\delta}}_{\mu}
 {\Gamma^{\epsilon\zeta}}_{\nu}\Sigma_{\gamma\delta}\Sigma_{\epsilon\zeta}\big).
\end{equation}
When calculating the commutator $ [\nabla_{\alpha},\nabla_{\beta}] $ we see that
$${e_{\alpha}}^{\mu}{e_{\beta}}^{\nu} \partial_{\mu}\partial_{\nu}- {e_{\beta}}^{\mu}{e_{\alpha}}^{\nu}
 \partial_{\mu}\partial_{\nu}={e_{\alpha}}^{\mu}{e_{\beta}}^{\nu} \partial_{\mu}\partial_{\nu}-
 {e_{\alpha}}^{\nu}{e_{\beta}}^{\mu}\partial_{\nu}\partial_{\mu}=0.$$
We have that
$$ {e_{\alpha}}^{\mu}{e_{\beta}}^{\nu}
\left(\frac{i}{2}{\Gamma^{\gamma\delta}}_{\nu}\Sigma_{\gamma\delta} \partial_{\mu}
 +\frac{i}{2}{\Gamma^{\gamma\delta}}_{\mu}\Sigma_{\gamma\delta}\partial_{\nu}\right)
 - {e_{\beta}}^{\mu}{e_{\alpha}}^{\nu}\left(
\frac{i}{2}{\Gamma^{\gamma\delta}}_{\nu}\Sigma_{\gamma\delta} \partial_{\mu}
 +\frac{i}{2}{\Gamma^{\gamma\delta}}_{\mu}\Sigma_{\gamma\delta}\partial_{\nu}\right)=0 .$$
Then the commutator becomes
\begin{eqnarray}
 [\nabla_{\alpha},\nabla_{\beta}] &=& {e_{\alpha}}^{\mu} {\omega^{\gamma}}_{\nu}( \partial_{\mu}{e_{\beta}}^{\nu} -
{e_{\delta}}^{\nu}\,{\Gamma^{\delta}}_{\mu\beta}) \nabla_{\gamma} -
{e_{\beta}}^{\mu} {\omega^{\gamma}}_{\nu}( \partial_{\mu}{e_{\alpha}}^{\nu} -
{e_{\delta}}^{\nu}\,{\Gamma^{\delta}}_{\mu\alpha}) \nabla_{\gamma}
 + {e_{\alpha}}^{\mu}{e_{\beta}}^{\nu}\left(\frac{i}{2}{\partial_{\mu}\Gamma^{\epsilon\zeta}}_{\nu}\Sigma_{\epsilon\zeta}
  -\frac{1}{4}{\Gamma^{\gamma\delta}}_{\mu}
 {\Gamma^{\epsilon\zeta}}_{\nu}\Sigma_{\gamma\delta}\Sigma_{\epsilon\zeta}\right)\cr
 &-&{e_{\beta}}^{\mu}{e_{\alpha}}^{\nu}\left(\frac{i}{2}{\partial_{\mu}\Gamma^{\epsilon\zeta}}_{\nu}\Sigma_{\epsilon\zeta}
  -\frac{1}{4}{\Gamma^{\gamma\delta}}_{\mu}
 {\Gamma^{\epsilon\zeta}}_{\nu}\Sigma_{\gamma\delta}\Sigma_{\epsilon\zeta}\right),\nonumber
\end{eqnarray}
and therefore
\begin{eqnarray}
 [\nabla_{\alpha},\nabla_{\beta}] &=&  [{\omega^{\gamma}}_{\nu}({e_{\alpha}}^{\mu}\partial_{\mu}{e_{\beta}}^{\nu} 
 - {e_{\beta}}^{\mu} \partial_{\mu}{e_{\alpha}}^{\nu})
 - {e_{\alpha}}^{\mu} {\omega^{\gamma}}_{\nu}{e_{\delta}}^{\nu}\,{\Gamma^{\delta}}_{\mu\beta} 
  +{e_{\beta}}^{\mu} {\omega^{\gamma}}_{\nu}{e_{\delta}}^{\nu}\,{\Gamma^{\delta}}_{\mu\alpha}]\nabla_{\gamma}\cr
 &+& {e_{\alpha}}^{\mu}{e_{\beta}}^{\nu}\left(\frac{i}{2}\partial_{\mu}{\Gamma^{\gamma\delta}}_{\nu}\Sigma_{\gamma\delta}
 -\frac{i}{2}\partial_{\nu}{\Gamma^{\gamma\delta}}_{\mu}\Sigma_{\gamma\delta}\right)
 - {e_{\alpha}}^{\mu}{e_{\beta}}^{\nu}\left(
  \frac{1}{4}{\Gamma^{\gamma\delta}}_{\mu}
 {\Gamma^{\epsilon\zeta}}_{\nu}\Sigma_{\gamma\delta}\Sigma_{\epsilon\zeta} 
  -\frac{1}{4}{\Gamma^{\gamma\delta}}_{\nu}
 {\Gamma^{\epsilon\zeta}}_{\mu}\Sigma_{\gamma\delta}\Sigma_{\epsilon\zeta}\right).\nonumber
\end{eqnarray}
Now using the identity  ${D_{\alpha\beta}}^{\gamma}= 
{\omega^{\gamma}}_{\nu}({e_{\alpha}}^{\mu}\partial_{\mu}{e_{\beta}}^{\nu} 
 - {e_{\beta}}^{\mu} \partial_{\mu}{e_{\alpha}}^{\nu})$ \cite{Nakahara, wytler} we have
\begin{eqnarray}
 [\nabla_{\alpha},\nabla_{\beta}] &=&  [{D_{\alpha\beta}}^{\gamma}
 - {\delta^{\gamma}}_{\delta}\,{\Gamma^{\delta}}_{\alpha\beta} 
  + {\delta^{\gamma}}_{\delta}\,{\Gamma^{\delta}}_{\beta\alpha}]\nabla_{\gamma}
 + {e_{\alpha}}^{\mu}{e_{\beta}}^{\nu}\left(\frac{i}{2}\partial_{\mu}{\Gamma^{\gamma\delta}}_{\nu}\Sigma_{\gamma\delta}
 -\frac{i}{2}\partial_{\nu}{\Gamma^{\gamma\delta}}_{\mu}\Sigma_{\gamma\delta}\right)
\cr
 &-&{e_{\alpha}}^{\mu}{e_{\beta}}^{\nu}\left(
  \frac{1}{4}{\Gamma^{\gamma\delta}}_{\mu}
 {\Gamma^{\epsilon\zeta}}_{\nu}\Sigma_{\gamma\delta}\Sigma_{\epsilon\zeta} 
  -\frac{1}{4}{\Gamma^{\gamma\delta}}_{\nu}
 {\Gamma^{\epsilon\zeta}}_{\mu}\Sigma_{\gamma\delta}\Sigma_{\epsilon\zeta}\right)\nonumber
\end{eqnarray}
or else
\begin{eqnarray}
 [\nabla_{\alpha},\nabla_{\beta}] &=&  -({\Gamma^{\gamma}}_{\alpha\beta} - {\Gamma^{\gamma}}_{\beta\alpha}
 - {D_{\alpha\beta}}^{\gamma}) \nabla_{\gamma}
 + {e_{\alpha}}^{\mu}{e_{\beta}}^{\nu}\left(\frac{i}{2}\partial_{\mu}{\Gamma^{\gamma\delta}}_{\nu}\Sigma_{\gamma\delta}
 -\frac{i}{2}\partial_{\nu}{\Gamma^{\gamma\delta}}_{\mu}\Sigma_{\gamma\delta}\right)
\cr
 &-&{e_{\alpha}}^{\mu}{e_{\beta}}^{\nu}\left(
  \frac{1}{4}{\Gamma^{\gamma\delta}}_{\mu}
 {\Gamma^{\epsilon\zeta}}_{\nu}\Sigma_{\gamma\delta}\Sigma_{\epsilon\zeta} 
  -\frac{1}{4}{\Gamma^{\gamma\delta}}_{\nu}
 {\Gamma^{\epsilon\zeta}}_{\mu}\Sigma_{\gamma\delta}\Sigma_{\epsilon\zeta}\right),\nonumber
\end{eqnarray}
the first term in parenthesis is the spacetime torsion \cite{Nakahara, wytler},
$$ {{\mathsf T}^{\gamma}}_{\alpha\beta} = {\Gamma^{\gamma}}_{\alpha\beta} - {\Gamma^{\gamma}}_{\beta\alpha}
 - {D_{\alpha\beta}}^{\gamma},$$
therefore we have
\begin{equation}
 [\nabla_{\alpha},\nabla_{\beta}] = - {{\mathsf T}^{\gamma}}_{\alpha\beta} \nabla_{\gamma}
+\frac{i}{2}{e_{\alpha}}^{\mu}{e_{\beta}}^{\nu}\left(\partial_{\mu}{\Gamma^{\gamma\delta}}_{\nu}
 -\partial_{\nu}{\Gamma^{\gamma\delta}}_{\mu}\right)\Sigma_{\gamma\delta}
 - \frac{1}{4}{e_{\alpha}}^{\mu}{e_{\beta}}^{\nu}{\Gamma^{\gamma\delta}}_{\mu}
 {\Gamma^{\epsilon\zeta}}_{\nu}\left(
  \Sigma_{\gamma\delta}\Sigma_{\epsilon\zeta}-
 \Sigma_{\epsilon\zeta}\Sigma_{\gamma\delta}\right).\nonumber
\end{equation}
Now we must use Lie algebra for operators $\Sigma_{\gamma\delta}$ 
in the tangent space, starting from equation (\ref{algebra_Lie}), where we have
\begin{equation}
\label{algebra_Lie_3}
 i[\Sigma_{\gamma\delta}, \Sigma_{\epsilon\zeta}] = \eta_{\epsilon\delta}\Sigma_{\gamma\zeta} -
  \eta_{\epsilon\gamma}\Sigma_{\delta\zeta} + \eta_{\zeta\delta}\Sigma_{\epsilon\gamma} -
  \eta_{\zeta\gamma}\Sigma_{\epsilon\delta},
\end{equation}
so that,
\begin{equation}
 [\nabla_{\alpha},\nabla_{\beta}] = - {{\mathsf T}^{\gamma}}_{\alpha\beta} \nabla_{\gamma}
+\frac{i}{2}{e_{\alpha}}^{\mu}{e_{\beta}}^{\nu}\left(\partial_{\mu}{\Gamma^{\gamma}}_{\delta\nu}
 -\partial_{\nu}{\Gamma^{\gamma}}_{\delta\mu}\right){\Sigma_{\gamma}}^{\delta}
 + \frac{i}{4}{e_{\alpha}}^{\mu}{e_{\beta}}^{\nu}{\Gamma^{\gamma\delta}}_{\mu}
 {\Gamma^{\epsilon\zeta}}_{\nu}\left(
  \eta_{\epsilon\delta}\Sigma_{\gamma\zeta} -
  \eta_{\epsilon\gamma}\Sigma_{\delta\zeta} + \eta_{\zeta\delta}\Sigma_{\epsilon\gamma} -
  \eta_{\zeta\gamma}\Sigma_{\epsilon\delta}\right).\nonumber
\end{equation}
Let us then simplify the last term of the above expression,
\begin{eqnarray}
 {\Gamma^{\gamma\delta}}_{\mu}
 {\Gamma^{\epsilon\zeta}}_{\nu}\left(
  \eta_{\epsilon\delta}\Sigma_{\gamma\zeta} -
  \eta_{\epsilon\gamma}\Sigma_{\delta\zeta} + \eta_{\zeta\delta}\Sigma_{\epsilon\gamma} -
  \eta_{\zeta\gamma}\Sigma_{\epsilon\delta}\right) &=& \cr
  {\Gamma^{\gamma}}_{\epsilon\mu}{\Gamma^{\epsilon}}_{\zeta\nu}{\Sigma_{\gamma}}^{\zeta}-
  (-{\Gamma^{\delta\gamma}}_{\mu}){\Gamma^{\epsilon}}_{\zeta\nu}\eta_{\epsilon\gamma}{\Sigma_{\delta}}^{\zeta}
  +(-{\Gamma^{\delta\gamma}}_{\mu}){\Gamma^{\epsilon\zeta}}_{\nu}\eta_{\zeta\delta}\Sigma_{\epsilon\gamma}-
  {\Gamma^{\gamma}}_{\delta\mu}{\Gamma^{\epsilon}}_{\gamma\nu}{\Sigma_{\epsilon}}^{\delta}&=&\cr
   {\Gamma^{\gamma}}_{\epsilon\mu}{\Gamma^{\epsilon}}_{\zeta\nu}{\Sigma_{\gamma}}^{\zeta}
  +{\Gamma^{\delta}}_{\epsilon\mu}{\Gamma^{\epsilon}}_{\zeta\nu}{\Sigma_{\delta}}^{\zeta}
  -{\Gamma^{\delta}}_{\gamma\mu}{\Gamma^{\epsilon}}_{\delta\nu}{\Sigma_{\epsilon}}^{\gamma}-
  {\Gamma^{\gamma}}_{\delta\mu}{\Gamma^{\epsilon}}_{\gamma\nu}{\Sigma_{\epsilon}}^{\delta}&=&\cr
  2 {\Gamma^{\gamma}}_{\delta\mu}{\Gamma^{\delta}}_{\epsilon\nu}{\Sigma_{\gamma}}^{\epsilon}
  -2 {\Gamma^{\delta}}_{\gamma\mu}{\Gamma^{\epsilon}}_{\delta\nu}{\Sigma_{\epsilon}}^{\gamma},\nonumber
\end{eqnarray}
that replacing for the commutator $ [\nabla_{\alpha},\nabla_{\beta}] $, it follows that
\begin{equation}
 [\nabla_{\alpha},\nabla_{\beta}] = - {{\mathsf T}^{\gamma}}_{\alpha\beta} \nabla_{\gamma}
+\frac{i}{2}{e_{\alpha}}^{\mu}{e_{\beta}}^{\nu}\left(\partial_{\mu}{\Gamma^{\gamma}}_{\delta\nu}
 -\partial_{\nu}{\Gamma^{\gamma}}_{\delta\mu}\right){\Sigma_{\gamma}}^{\delta}
 +\frac{i}{2}{e_{\alpha}}^{\mu}{e_{\beta}}^{\nu}
 (  {\Gamma^{\gamma}}_{\delta\mu}{\Gamma^{\delta}}_{\epsilon\nu}{\Sigma_{\gamma}}^{\epsilon}
  - {\Gamma^{\delta}}_{\epsilon\mu}{\Gamma^{\gamma}}_{\delta\nu}{\Sigma_{\gamma}}^{\epsilon}),\nonumber
\end{equation}
or
\begin{equation}
  [\nabla_{\alpha},\nabla_{\beta}] =  - {{\mathsf T}^{\gamma}}_{\alpha\beta} \nabla_{\gamma}
 +\frac{i}{2}{e_{\alpha}}^{\mu}{e_{\beta}}^{\nu}\left(\partial_{\mu}{\Gamma^{\gamma}}_{\delta\nu}
 -\partial_{\nu}{\Gamma^{\gamma}}_{\delta\mu}
  + {\Gamma^{\gamma}}_{\epsilon\mu}{\Gamma^{\epsilon}}_{\delta\nu} 
  - {\Gamma^{\epsilon}}_{\delta\mu}{\Gamma^{\gamma}}_{\epsilon\nu}
 \right){\Sigma_{\gamma}}^{\delta}.
\end{equation}
The term in parenthesis is the curvature tensor where, 
\begin{equation}
 {R^{\gamma}}_{\delta\mu\nu}= {\partial_{\mu}\Gamma^{\gamma}}_{\delta\nu}
 -{\partial_{\nu}\Gamma^{\gamma}}_{\delta\mu}
  + {\Gamma^{\gamma}}_{\epsilon\mu}{\Gamma^{\epsilon}}_{\delta\nu} 
  - {\Gamma^{\epsilon}}_{\delta\mu}{\Gamma^{\gamma}}_{\epsilon\nu}.
\end{equation}
So finally we have
\begin{equation}
 \label{Operador_curvatura_9}
  [\nabla_{\alpha},\nabla_{\beta}] =  - {{\mathsf T}^{\gamma}}_{\alpha\beta} \nabla_{\gamma}
 +\frac{i}{2}{e_{\alpha}}^{\mu}{e_{\beta}}^{\nu}{R^{\gamma}}_{\delta\mu\nu} {\Sigma_{\gamma}}^{\delta},
\end{equation}
or else
\begin{equation}
 \label{Operador_curvatura_10}
  [\nabla_{\alpha},\nabla_{\beta}] =  - {{\mathsf T}^{\gamma}}_{\alpha\beta} \nabla_{\gamma}
 +\frac{i}{2}\,{R^{\gamma}}_{\delta\alpha\beta} {\Sigma_{\gamma}}^{\delta}.
\end{equation}
If we use the above operation in a vector $V^{\epsilon}$, we must use the generator ${\Sigma_{\gamma}}^{\delta}$ 
applied to the vector, so that we have
\begin{equation}
 {\Sigma_{\gamma}}^{\delta}V^{\epsilon} = i(\eta^{\delta\epsilon}\eta_{\gamma\zeta}-{\delta_{\gamma}}^{\epsilon}{\delta_{\zeta}}^{\delta})V^{\zeta},\nonumber
\end{equation}
and replacing in equation (\ref{Operador_curvatura_10}) we get exactly the same value from the equation (\ref{comutador_derivadas_covariantes}).

%\bibliographystyle{elsarticle-num}

\end{document}